\documentclass[preprint,showpacs,preprintnumbers,amsmath,amssymb,nofootinbib]{revtex4}

\usepackage{graphicx}
\usepackage{dcolumn}
\usepackage{bm}

\begin{document}
\title{Self-consistent quasiparticle RPA for multi-level pairing model}
\author{N. Quang Hung$^{1}$}
 \altaffiliation[On leave of absence from the ]{Institute of Physics and Electronics, Hanoi, Vietnam}
  \email{nqhung@riken.jp}
\author{N. Dinh Dang$^{1,2}$}%
 \email{dang@riken.jp}
\affiliation{$^{1}$ Heavy-Ion Nuclear Physics Laboratory, RIKEN Nishina Center
for Accelerator-Based Science, 
2-1 Hirosawa, Wako City, 351-0198 Saitama, Japan\\
$^{2}$ Institute for Nuclear Science and Technique, Vietnam Atomic
Energy Commission, Hanoi, Vietnam}
\date{\today}
\begin{abstract}
Particle-number projection within the Lipkin-Nogami (LN) method is
applied to the self-consistent quasiparticle random-phase approximation
(SCQRPA), which is tested in an exactly solvable multi-level pairing model. The SCQRPA
equations are numerically solved to find the energies of the
ground and excited states at various numbers $\Omega$ of
doubly degenerate equidistant levels. The use of the LN
method allows one to avoid the collapse of the BCS (QRPA) to obtain the
energies of the ground and excited states as smooth functions
of the interaction parameter $G$. The comparison between results given
by different approximations such as the SCRPA, QRPA, LNQRPA, SCQRPA
and LNSCQRPA is carried out. While the use of the LN method
significantly improves the agreement with the exact results 
in the
intermediate coupling region,
we
found that in the strong coupling region the SCQRPA results are 
closest to the exact ones.
\end{abstract}

\maketitle
\section{\label{sec1} INTRODUCTION}
\indent The random-phase approximation (RPA), which includes
correlations in the ground state, provides a simple theory of
excited states of the nucleus. However, the RPA breaks down at a
certain value $G_{\rm cr}$ of interaction parameter $G$, where it yields imaginary
eigenvalues. The reason is that the RPA equations, linear with
respect to the $X$ and $Y$ amplitudes of the RPA excitation
operator, are derived based on the quasi-boson approximation (QBA).
The latter neglects the Pauli principle between fermion pairs and
its validity is getting poor with increasing the interaction
parameter $G$. The collapse of the RPA at the critical value 
$G_{\rm cr}$ of $G$ invalidates the use of the QBA.
The RPA therefore needs to be extended to correct this deficiency, 
at 
least for finite systems such as nuclei.
\\
\indent One of methods to restore the Pauli principle is to
renormalize the conventional RPA to include the non-zero values of
the commutator between the fermion-pair operators in the correlated
ground state. These so-called ground-state correlations beyond RPA are
neglected within the QBA. The interaction in this way is
renormalized and the collapse of RPA is avoided. The resulting
theory is called the renormalized RPA (RRPA)
\cite{RRPA1,Rowe,RRPA5}. However, the test of the RRPA
carried out within several exactly solvable models showed that the
RRPA results are still far from the exact solutions
\cite{RRPA5,SCRPA1,SCRPA2}.\\
\indent Recently, a significant development in improving the RPA has
been carried out within the self-consistent RPA (SCRPA)
\cite{SCRPA1,SCRPA2,SCRPA3}. Based on the same concept of renormalizing the
particle-particle ($pp$) RPA, the SCRPA made a step forward by 
including the screening factors, which are
the expectation values of the products of two pairing operators in
the correlated ground state. The SCRPA has been applied to the
exactly solvable multi-level pairing model, where the energies of
the ground state and first excited state in the system with $N+2$
particles relative to the energy of the ground-state level in the $N$-particle
system are calculated and compared with the exact results. It has
been found that the agreement with the exact solutions is good only
in the weak coupling region, where the pairing-interaction parameter
$G$ is smaller than the critical values $G_{\rm cr}$. 
In the strong coupling
region ($G>>G_{\rm cr}$), the agreement between the SCRPA and exact
results becomes poor~\cite{SCRPA1,SCRPA2}. In this region a quasiparticle representation
should be used in place of the $pp$ one, as has been pointed out in Ref. \cite{DaTa}. 
As a matter of fact, an
extended version of the SCRPA in the superfluid region has been
proposed and is called the self-consistent quasiparticle RPA
(SCQRPA), which was applied for the first time to the seniority
model in Ref. \cite{SCQRPA1} and a two-level pairing model in Ref.
\cite{SCQRPA2}. However, the SCQRPA also 
collapses at $G=G_{\rm cr}$. It is therefore highly desirable to
develop a SCQRPA that works at all values of $G$ and also in more realistic
cases, e.g. multi-level models. The aim of the present work is to
construct such an approach. Obviously, the collapse of the SCQRPA at
$G=G_{\rm cr}$, which is the same as that of the
non-trivial solution for the pairing gap within the Bardeen-Cooper-Schrieffer theory
(BCS), can
be removed by performing the particle-number projection (PNP). The
Lipkin-Nogami method \cite{Lipkin,Nogami5}, 
which is an approximated PNP before variation,
will be used in such extension of the SCQRPA in the present paper
because of its simplicity. This approach shall be applied to a
multi-level pairing model, the so-called Richardson model
\cite{Richardson-model}, which is an exactly solvable model
extensively employed in literature to test approximations of
many-body problems.\\
\indent The paper is organized as follows. Section \ref{sec2} presents a
brief outline of the SCQRPA theory that includes the PNP within the LN method. The results of
numerical calculations are analyzed and  discussed in sec. \ref{sec3}. Conclusions are
drawn in the last section.
\section{\label{sec2}FORMALISM}
\subsection{\label{subsec2.1}Model Hamiltonian}
The Richardson model (also called the multi-level pairing model, 
picket-fence model or ladder model, etc) was described in detail in
Refs. \cite{Richardson-model,SCRPA1,SCRPA2,SCRPA3}. It consists of
$\Omega$ doubly-fold equidistant levels interacting via a pairing force
with a constant parameter $G$. The model Hamiltonian is given as
\begin{equation}\label{eq1}
    H=\sum_{j=1}^\Omega(\epsilon_j-\lambda)N_j-G\sum_{j,j'=1}^{\Omega}{P_j^{\dagger}P_{j'}}~,
\end{equation}
where $\epsilon_j$ are the single-particle energies on the j-shells.
The particle-number operator $N_j$ and pairing operators $P_j^{\dagger}$,
$P_j$ on the $j$-th orbital (with unit shell degeneracy $j+1/2\equiv
1$) are defined as
\begin{eqnarray}
&&N_j=a_j^{\dagger}a_j+a_{-j}^{\dagger}a_{-j}~,\label{eq2.1}\\
&&P_j^{\dagger}=a_j^{\dagger}a_{-j}^{\dagger}~, \hspace{5mm}
P_j=(P_j^{\dagger})^{\dagger}~. \label{eq2.2}
\end{eqnarray}
These operators fulfill the following exact commutation relations
\begin{eqnarray}
&&[P_j~,~ P_{j'}^{\dagger}]=\delta_{jj'}(1-N_j)~, \label{eq3.1}\\
&&[N_j~,~ P_{j'}^{\dagger}]=2\delta_{jj'}P_{j'}^{\dagger}~, \hspace{5mm}
[N_j~,~ P_{j'}]=-2\delta_{jj'}P_{j'}~. \label{eq3.2}
\end{eqnarray}
By using the Bogoliubov transformation from particle operators
$a_j^{\dagger}$ and $a_j$ 
to quasiparticle ones $\alpha_j^{\dagger}$ and $\alpha_j$
\begin{equation}\label{eq4}
a_j^{\dagger}=u_j\alpha_j^{\dagger} + v_j\alpha_{-j}~, \hspace{5mm}
a_{-j}=u_j\alpha_{-j}-v_j\alpha_{j}^{\dagger}~,
\end{equation}
the pairing Hamiltonian in Eq. \eqref{eq1} is transformed into the
quasiparticle Hamiltonian as \cite{quasi-ha1,quasi-ha2}
\[
H=a+\sum_j{b_j\mathcal{N}_j}+\sum_j{c_j(\mathcal{A}_j^{\dagger}+\mathcal{A}_j})
+\sum_{jj'}{d_{jj'}\mathcal{A}_j^{\dagger}\mathcal{A}_{j'}}
+\sum_{jj'}{g_j(j')(\mathcal{A}_{j'}^{\dagger}\mathcal{N}_j+\mathcal{N}_j\mathcal{A}_{j'})}
\]
\begin{equation}\label{eq5}
+\sum_{jj'}{h_{jj'}(\mathcal{A}_j^{\dagger}\mathcal{A}_{j'}^{\dagger}
+\mathcal{A}_{j'}\mathcal{A}_j)}+\sum_{jj'}{q_{jj'}\mathcal{N}_j\mathcal{N}_{j'}}~,
\end{equation}
where $\mathcal{N}_{j}$ is the quasiparticle-number operator, while 
$\mathcal{A}_j^{\dagger}$ and $\mathcal{A}_j$ are the creation and
destruction operators of a pair of time-conjugated quasiparticles:
\begin{eqnarray}
&&\mathcal{N}_j=\alpha_j^{\dagger}\alpha_j+\alpha_{-j}^{\dagger}\alpha_{-j}\label{eq6.1}~,\\
&&\mathcal{A}_j^{\dagger}=\alpha_j^{\dagger}\alpha_{-j}^{\dagger}~, \hspace{5mm}
\mathcal{A}_j=(\mathcal{A}_j^{\dagger})^{\dagger}~.\label{eq6.2}
\end{eqnarray}
The commutation relations between operators $\mathcal{N}_{j}$, $\mathcal{A}_j^{\dagger}$ and
$\mathcal{A}_j$ are similar to those for particle operators in Eqs.
\eqref{eq3.1} and \eqref{eq3.2}, namely
\begin{eqnarray}
&&[\mathcal{A}_j~,~\mathcal{A}_{j'}^{\dagger}]=\delta_{jj'}(1-\mathcal{N}_j)~,\label{eq7.1}\\
&&[\mathcal{N}_j~,~\mathcal{A}_{j'}^{\dagger}]=2\delta_{jj'}\mathcal{A}_{j'}^{\dagger}~,
\hspace{5mm}
[\mathcal{N}_j~,~\mathcal{A}_{j'}]=-2\delta_{jj'}\mathcal{A}_{j'}~.\label{eq7.2}
\end{eqnarray}
The coefficients $a$, $b_j$, $c_j$, $d_{jj'}$, $g_j(j')$, $h_{jj'}$, $q_{jj'}$
in Eq. \eqref{eq5} are given in terms of the coefficients $u_j$, $v_j$
of the
Bogoliubov transformation, and the single particle
energies $\epsilon_j$ as (see, e.g. Ref. \cite{quasi-ha1,quasi-ha2})
\begin{equation}
a=2\sum_j{(\epsilon_j-\lambda)v_j^2}-G\left(\sum_j{u_jv_j}\right)^2-G\sum_j{v_j^4}~,\label{eq8.1}
\end{equation}
\begin{equation}
b_j=(\epsilon_j-\lambda)(u_j^2-v_j^2)+2Gu_jv_j\sum_{j'}{u_{j'}v_{j'}}+Gv_j^4~,\label{eq8.2}
\end{equation}
\begin{equation}
c_j=2(\epsilon_j-\lambda)u_jv_j-G(u_j^2-v_j^2)\sum_{j'}{u_{j'}v_{j'}}-2Gu_jv_j^3~,\label{eq8.4}
\end{equation}
\begin{equation}
d_{jj'}=-G(u_j^2u_{j'}^2+v_j^2v_{j'}^2)=d_{j'j}~,\label{eq8.5}
\end{equation}
\begin{equation}
g_{j}(j')=Gu_jv_j(u_{j'}^2-v_{j'}^2)~,\label{eq8.6}
\end{equation}
\begin{equation}
h_{jj'}=\frac{G}{2}(u_j^2v_{j'}^2+v_{j}^2u_{j'}^2)=h_{j'j}~,\label{eq8.7}
\end{equation}
\begin{equation}
q_{jj'}=-Gu_jv_ju_{j'}v_{j'}=q_{j'j}~.\label{eq8.8}
\end{equation}
The single-particle energies are given as $\epsilon_j=j\epsilon$, where
$j=$ 1, $\dots$, $\Omega$, and $\epsilon$ is the level distance chosen
to be equal to 1 MeV in the present work. The chemical potential $\lambda$ and the coefficients $u_j$ and
$v_j$ are determined by solving the
gap equations discussed in the next section.
\subsection{\label{subsec2.2}Gap and number equations}
\subsubsection{\label{subsec2.2.1} Renormalized BCS}
It is well known that the Pauli principle between the
quasiparticle-pair operators $\mathcal{A}_j$ and
$\mathcal{A}_{j'}^{\dagger}$ is neglected within the
conventional BCS, which assumes that $\langle{\rm
BCS}|{\cal N}_{j}|{\rm BCS}\rangle=$ 0 within the BCS ground state $|{\rm BCS}\rangle$. 
A simple way to restore
the Pauli principle is to introduce a new ground state
$|\bar{0}\rangle$ in which the correlations among quasiparticles
lead to nonzero values of the quasiparticle occupation numbers so that
the contribution of the $\mathcal{N}_{j}$-term at the right-hand side (rhs) of Eq.
(\ref{eq7.1}) is preserved. By
doing so, the BCS equations are renormalized and the resulting
theory is called the renormalized BCS (RBCS) \cite{RBCS}. 
Within the
RBCS the commutator between the quasiparticle-pair operators are defined
as
\begin{equation}\label{eq14}
\langle\bar{0}|[\mathcal{A}_j,\mathcal{A}_{j'}^{\dagger}]|\bar{0}\rangle
 =\delta_{jj'}\langle{\cal D}_{j}\rangle~,
\end{equation}
with 
\begin{equation}
{\cal D}_{j}=1-{\cal N}_{j}~,\hspace{5mm} 
\langle{\cal D}_{j}\rangle=1-2n_j~,
\label{Dj}
\end{equation}
where $n_j$ is the quasiparticle number in the
correlated ground state $|\bar{0}\rangle$
\begin{equation}\label{eq15}
n_j\equiv\frac{1}{2}\langle\bar{0}|\mathcal{N}_j|\bar{0}\rangle \neq
0~.
\end{equation}
Taking into account Eq. \eqref{eq14} and performing a constrained
variational calculation to minimize the Hamiltonian
$H\equiv H'-\lambda\hat{N}$, where $\hat{N}=\sum_jN_j$ is the particle-number
operator, the RBCS equations for the pairing gap $\Delta$ and particle
number $N$ have been derived as~\cite{RBCS} 
\begin{equation}\label{eq16}
{\Delta}=G\sum_j\tau_j~,\hspace {2mm} N=2\sum_j\rho_j~,
\end{equation}
where
\begin{equation}
\tau_j=u_jv_j\langle{\cal D}_{j}\rangle~,\hspace{5mm}
\rho_j=v_j^2\langle{\cal D}_{j}\rangle+\frac{1}{2}(1-\langle{\cal D}_{j}\rangle)~,
\label{eq16.1}
\end{equation}
\begin{equation}
u_j^2=\frac{1}{2}\left(1+\frac{\epsilon_j-Gv_j^2-\lambda}{E_j}\right)~, 
\hspace{5mm}
v_j^2=\frac{1}{2}\left(1-\frac{\epsilon_j-Gv_j^2-\lambda}{E_j}\right)~,
\label{uv}
\end{equation}
\begin{equation}
{E}_j=\sqrt{(\epsilon_j-Gv_j^2-\lambda)^2+{\Delta}^2}
\label{eq16.2}~.
\end{equation}
The renormalization factors $\langle{\cal D}_{j}\rangle$, called the ground-state correlation
factors, are obtained by solving the SCQRPA
equations discussed later in this paper (See Sec. \ref{subsec2.3.2}). The internal energy of 
the system within the RBCS ground state (the RBCS ground-state energy)
is given as
\begin{equation}\label{eq16.4}
E_{\rm g.s.}^{\rm RBCS}=2\sum_j(\epsilon_j-\lambda){\rho}_{j}-\frac{{\Delta}^2}{G}-G\sum_j
{\rho}_j^2~.
\end{equation}
By setting $\langle{\cal D}_{j}\rangle = 1$, the RBCS equations go back to the
well-known BCS ones.
\subsubsection{\label{subsec2.2.2}BCS with SCQRPA correlations}
In the minimization procedure, which leads to the equation (See, e.g.
Ref. ~\cite{DukSchu})
\begin{equation}
    \langle\bar{0}|[H,{\cal A}_{j}^{\dagger}]|\bar{0}\rangle=0~,
    \label{mini}
    \end{equation}
the RBCS ignores the expectation values 
$\langle\mathcal{A}_{j'}^{\dagger}\mathcal{A}_j\rangle\equiv
\langle\bar{0}|\mathcal{A}_{j'}^{\dagger}\mathcal{A}_j|\bar{0}\rangle$ and
$\langle\mathcal{A}_{j'}\mathcal{A}_j\rangle\equiv
\langle\bar{0}|\mathcal{A}_{j'}\mathcal{A}_j|\bar{0}\rangle$ of the
products of pair operators
in the correlated
quasiparticle ground state $|\bar 0\rangle$. By retaining these screening
factors in calculating the left-hand side (lhs) of Eq. (\ref{mini}), 
we derive from Eq. (\ref{mini}) an equation for the level-dependent 
pairing gap in the form
\begin{equation}
\Delta_{j}=G\frac{\sum_{j'}u_{j'}v_{j'}\langle{\cal D}_{j}{\cal D}_{j'}\rangle}
{\langle{\cal D}_{j}\rangle}~,
\label{scgap}
\end{equation}
with the single-particle energies $\epsilon_{j}$ in the expressions for
$u_{j}$ and $v_{j}$ in Eq. (\ref{uv}) being renormalized to
$\epsilon_{j}'$ as
\begin{equation}
    \epsilon_{j}'=\epsilon_{j}+\frac{G}{\langle{\cal D}_{j}\rangle}
    \sum_{j'}(u_{j'}^{2}-v_{j'}^{2})\bigg
    (\langle{\cal A}_{j}^{\dagger}{\cal A}_{j'}^{\dagger}\rangle
    +\langle{\cal A}_{j}^{\dagger}{\cal A}_{j'}\rangle\bigg)~.
    \label{reeps}
    \end{equation}
We call Eq. (\ref{scgap}) the BCS gap equation with SCQRPA
correlations, and use the abbreviation BCS1 to denote this 
approach, having in mind that it includes the 
screening factors $\langle{\cal A}_{j}^{\dagger}{\cal
A}_{j'}^{\dagger}\rangle$ and 
$\langle{\cal A}_{j}^{\dagger}{\cal A}_{j'}\rangle$ in the
renormalized single-particle energies given by Eq.
(\ref{reeps}). These screening factors 
are found by solving Eqs. (\ref{scgap}) and (\ref{reeps}) selfconsistently with 
the SCQRPA ones to be discussed later in Sec. \ref{subsec2.3},
where the explicit expressions of the screening factors are given 
in terms of the SCQRPA forward- and backward-going 
(${\cal X}$ and ${\cal Y}$) amplitudes. The limit 
case of Eqs. (\ref{scgap}) and (\ref{reeps}) for a degenerate
two-level model is studied in Ref. \cite{SCQRPA2}. 

The rhs of Eq. (\ref{scgap})
contains the expectation values $\langle{\cal D}_{j}
{\cal D}_{j'}\rangle$, whose exact treatment is not possible
as it involves an infinite series in 
terms of the products of 
${\cal A}_{j}^{\dagger}{\cal A}_{j}{\cal A}_{j'}^{\dagger}{\cal A}_{j'}$
~\cite{SCQRPA2}, or an infinite boson expansion~\cite{Samba}, which
again needs to be truncated at a certain order. In Ref.
\cite{SCQRPA2} this series is truncated at the first order, while the
consideration in Ref.
\cite{Samba} is limited up to the four-boson
terms. Such expansion is
based on the method of treating the single-particle (quasiparticle) 
density used by Rowe in Ref. \cite{Rowe} or a mapping employed in Ref.
\cite{RRPA5}. In the numerical calculations within the 
present paper we treat these terms approximately as follows. By noticing that 
the expectation values 
    $\langle{\cal D}_{j}{\cal D}_{j'}\rangle$ are present in 
in the ratios $\langle{\cal D}_{j}{\cal D}_{j'}\rangle/
    {\langle{\cal D}_{j}\rangle}$
    or, more general, $\langle{\cal D}_{j}{\cal D}_{j'}\rangle/
    \sqrt{\langle{\cal D}_{j}\rangle\langle{\cal D}_{j'}\rangle}$, and that
    \begin{equation}
    \langle{\cal D}_{j}{\cal D}_{j'}\rangle=\langle{\cal D}_{j}\rangle
    \langle{\cal D}_{j'}\rangle + \delta{\cal N}_{jj'}~,
    \hspace{5mm} {\rm with}\hspace{5mm}\delta{\cal N}_{jj'}=
    \langle{\cal N}_{j}{\cal
    N}_{j'}\rangle - \langle{\cal N}_{j}\rangle\langle{\cal
    N}_{j'}\rangle~,
    \label{DD}
    \end{equation}    
we rewrite these ratios as
\begin{equation}
    \frac{\langle{\cal D}_{j}{\cal D}_{j'}\rangle}{
        \sqrt{\langle{\cal D}_{j}\rangle\langle{\cal D}_{j'}\rangle}}
        = \sqrt{\langle{\cal D}_{j}\rangle\langle{\cal D}_{j'}\rangle} +
        \frac{\delta{\cal N}_{jj'}}
        {\sqrt{\langle{\cal D}_{j}\rangle\langle{\cal D}_{j'}\rangle}}~.
        \label{ratio}
        \end{equation}
        The numerator $\delta{\cal N}_{jj'}$ of the last term at
        the rhs of Eq. (\ref{ratio}) can be estimated by using the
        mean-field contraction as
\begin{equation}
    \delta{\cal N}_{jj'}\simeq
    2\delta_{jj'}n_{j}(1-n_{j}) = \delta_{jj'}(\delta{\cal N}_{j})^2~,
    \label{deltaN}
        \end{equation}
where $(\delta{\cal N}_{j})^2\equiv\langle{\cal N}_{j}^{2}\rangle-
\langle{\cal N}_{j}\rangle^{2}=2n_{j}(1-n_{j})$ is the
quasiparticle-number fluctuation on the $j$-th orbital. This quantity 
is much smaller than 1, while the denominator 
$\sqrt{\langle{\cal D}_{j}\rangle\langle{\cal D}_{j'}\rangle}$, which 
is also the first term at the rhs of Eq. (\ref{ratio}), is
comparable with 1 as the ground-state correlations factors 
$\langle{\cal D}_{j}\rangle$ are not much smaller than 1. 
Therefore the last term at the rhs of Eq. (\ref{ratio}) can be safely 
neglected so that 
\begin{equation}
    \frac{\langle{\cal D}_{j}{\cal D}_{j'}\rangle}{
        \sqrt{\langle{\cal D}_{j}\rangle\langle{\cal D}_{j'}\rangle}}
        \simeq \sqrt{\langle{\cal D}_{j}\rangle\langle{\cal
        D}_{j'}\rangle}~.
        \label{approx1}
        \end{equation}
        Consequently, the ratio $\langle{\cal D}_{j}{\cal D}_{j'}\rangle/
        \langle{\cal D}_{j}\rangle$ in the sum over $j'$ at the rhs of
        Eq. (\ref{scgap})
        can be simply approximated with $\langle {\cal
        D}_{j'}\rangle$~\footnote{In Refs. 
\cite{SCRPA1,SCRPA2} the factorization $\langle{N}_{j}{N}_{j'}\rangle\simeq
\langle{N}_{j}\rangle\langle{N}_{j'}\rangle$ ($j=$ $p$, $h$) 
was straightforwardly used to close the SCRPA equations because 
$\langle N_{h}\rangle\langle N_{h'}\rangle$, whose value in
the Hartree-Fock (HF) limit is 4, is much
larger than the particle-number fluctuation 
$(\delta N_{h})^2=2f_{h}(1-f_{h})$. This is no longer the case for
quasiparticle numbers, where $(\delta{\cal N}_{j})^{2}$ are
of the same order with $\langle{\cal N}_{j}\rangle^{2}$ .}. 
In this case Eq. (\ref{scgap}) takes the same
        level-independent form as 
        that of Eq. (\ref{eq16}) for the RBCS gap except that the
        single-particle energies in $u_{j'}$ and $v_{j'}$ are now
        given by Eq. (\ref{reeps}). In the rest of the paper, such
        level-independent approximation for the pairing gap is
        assumed, whose numerical accuracy is checked in the Appendix
        \ref{appendix}.    
\subsection{\label{subsec2.2.3}Lipkin-Nogami method with SCQRPA
correlations}

The main drawback of the BCS is that its wave
function is not an eigenstate of the particle-number operator
$\hat{N}$. The BCS, 
therefore, suffers from an inaccuracy caused by the particle-number
fluctuations. The collapse of the BCS at a
critical value $G_{\rm cr}$ of the pairing parameter $G$, below which it
has only a trivial solution with zero pairing gap, is intimately
related to the particle-number fluctuations within BCS~\cite{Nogami5}.
This defect is cured by projecting out the component of the
wave-function that corresponds to the right number of particles. The 
Lipkin-Nogami (LN) method is an approximated PNP, which has been
shown to be simple and yet efficient in many realistic
calculations 
(See Ref. \cite{Egido} for a recent detailed clarification of the use 
of the LN method).
This method, discussed in detail in Refs.
\cite{Lipkin,Nogami5}, 
is a PNP before variation based 
on the BCS wave function, therefore
the Pauli principle between the quasiparticle-pair operators Eq.
(\ref{eq7.1}) 
is still neglected within the original version of this method. In the
present work, to restore the Pauli principle 
we propose a renormalization of the LN method, which we refer to as the 
renormalized LN (RLN) method or LN method with SQRPA correlations (LN1)  
when they are based on the RBCS 
or BCS1, respectively. 
Similar to the BCS1 (RBCS), the LN1 (RLN) includes the
quasiparticle correlations in the correlated ground state
$|\bar{0}\rangle$, and the LN1 (RLN) equations are obtained by carrying out
the variational calculation to minimize Hamiltonian
$\tilde{H}\equiv H'-\lambda\hat{N}-\lambda_2\hat{N}^2$. The LN1 
equations obtained in
this way have the form
\begin{equation}
\tilde\Delta=G\sum_j\tilde\tau_j~,\hspace{5mm}
N=2\sum_j\tilde\rho_j~,\label{eq17.1}
\end{equation}
\begin{equation}
\tilde\epsilon_{j}=\epsilon'_j+(4\lambda_2-G){\tilde{v}_j}^2~,\hspace{5mm}
\lambda=\lambda_1+2\lambda_2(N+1)~,\label{eq17.2}
\end{equation}
where
\begin{equation}
\tilde{\tau}_j=\tilde{u}_j\tilde{v}_j\langle{\cal D}_{j}\rangle~,
\hspace{5mm} \tilde{\rho}_j=\tilde{v}_j^2\langle{\cal D}_{j}\rangle+\frac{1}{2}(1-\langle{\cal D}_{j}\rangle)~,\label{rhotau}
\end{equation}
\begin{equation}
    \tilde{u}_j^2=\frac{1}{2}\left(1+\frac{\tilde{\epsilon}_j
    -\lambda}{\tilde{E}_j}\right)~, 
\hspace{5mm}
\tilde{v}_j^2=\frac{1}{2}\left(1-\frac{\tilde{\epsilon}_j
-\lambda}{\tilde{E}_j}\right)~,
\hspace{5mm}
\tilde{E}_j=\sqrt{(\tilde{\epsilon}_j-\lambda)^2+\tilde\Delta^2}\label{u'v'}~.
\end{equation}
The coefficient
$\lambda_2$ has the following form~\cite{Lambda2}
\begin{equation}\label{eq18}
\lambda_{2}=\frac{G}{4}
\frac{\sum_j(1-\tilde\rho_j)\tilde\tau_j
\sum_{j'}\tilde\rho_{j'}\tilde\tau_{j'}-
\sum_j(1-\tilde\rho_j)^2\tilde\rho_j^2}
{\left[\sum_j\tilde\rho_j(1-\tilde\rho_j)\right]^2
-\sum_j(1-\tilde\rho_j)^2\tilde\rho_j^2}~,
\end{equation}
which becomes the expression given in the original paper \cite{Nogami5}
of the LN method
in the limit of $\langle{\cal D}_{j}\rangle=$ 1 and
$\epsilon_{j}'=\epsilon_{j}$.
The internal energy obtained within the LN1 ground state (the LN1
ground-state energy) is given as
\begin{equation}\label{eq19}
E_{\rm g.s.}^{\rm LN1}=2\sum_j{(\epsilon_j-\lambda)\tilde{\rho}_j}-
\frac{\tilde\Delta^2}{G}-G\sum_j
\tilde{\rho}_j^2-\lambda_2\Delta
N^2~,
\end{equation}
where the expression for the 
particle-number fluctuation $\Delta N^2$ in terms of $\tilde{u}_{j}$, 
$\tilde{v}_{j}$ and $n_{j}\equiv(1-\langle{\cal D}_{j}\rangle)/2$ has been derived in
Ref. \cite{quasi-ha2}.
The LN1 equations becomes the RLN equations by replacing
the renormalized single-particle energies 
$\epsilon'_{j}$ defined in Eq. (\ref{reeps}) with $\epsilon_{j}$. 
The RLN equations return to the BCS ones in the limit case, when
$\lambda_2=0$ and $\langle{\cal D}_{j}\rangle=1$.
\subsection{\label{subsec2.3}SCQRPA equations}
\subsubsection{\label{subsec2.3.1}QRPA}
The QRPA excited state $|\nu\rangle$ is constructed by acting
the QRPA operator $Q_\nu^{\dagger}$ 
\begin{equation}\label{eq20}
Q_\nu^{\dagger}=\sum_j(X_j^\nu\mathcal{A}_j^{\dagger}-Y_j^\nu\mathcal{A}_j)~,
\hspace{5mm} Q_\nu = (Q_\nu^{\dagger})^{\dagger}~,
\end{equation}
on the QRPA ground state $|0\rangle$ as
\begin{equation}\label{eq21}
|\nu\rangle=Q_\nu^{\dagger}|0\rangle~,
\end{equation}
where $|0\rangle$ is defined as the vacuum for the operator
\eqref{eq20}, i.e.
\begin{equation}\label{eq22}
Q_\nu|0\rangle=0~.
\end{equation}
The QBA assumes the following relation
\begin{equation}\label{eq23}
\langle
0|[\mathcal{A}_j,\mathcal{A}_{j'}^{\dagger}]|0\rangle=\delta_{jj'}~.
\end{equation}
Within the QBA the QRPA amplitude $X_j^{\nu}$ and $Y_j^{\nu}$ 
obey the well-known normalization
(orthogonality) conditions
\begin{equation}\label{eq24}
\sum_j\left(X_j^\nu X_j^{\nu'}-Y_j^\nu
Y_j^{\nu'}\right)=\delta_{\nu\nu'}~,
\end{equation}
to guarantee that the QRPA operators \eqref{eq20} are bosons, i.e.
\begin{equation}\label{eq25}
\langle0|\left[Q_\nu ,
Q_{\nu'}^{\dagger}\right]|0\rangle=\delta_{\nu\nu'}~.
\end{equation}
By linearizing the equation of motion with respect to Hamiltonian
\eqref{eq5} and operators \eqref{eq20}, the set of linear QRPA
equations is derived and presented in the matrix form as follow
\begin{equation}\label{eq26}
\left(\begin{array}{cc}A&B\\B&A\end{array}\right)
\left(\begin{array}{cc}X_j^\nu\\Y_j^\nu\end{array}\right)
=\omega_\nu\left(\begin{array}{cc}X_j^\nu\\-Y_j^\nu\end{array}\right)~,
\end{equation}
where the QRPA submatrices are given as
\begin{eqnarray}
A_{jj'}&=&2(b_j+2q_{jj'})\delta_{jj'}+d_{jj'}~,\label{eq27.1}\\
B_{jj'}&=&2(1-\delta_{jj'})h_{jj'}\label{eq27.2}~,
\end{eqnarray}
and the eigenvalues $\omega_{\nu}\equiv{\cal E}_{\nu}-{\cal E}_{0}$
are the energies ${\cal E}_{\nu}$ of the excited states relative to 
that of the ground-state level, ${\cal E}_{0}$. 
The QRPA ground-state energy is
given as the sum of the BCS ground-state energy $E_{\rm g.s.}^{\rm BCS}$ 
and the QRPA correlation
energy as follows~\cite{Rowe,QRPA1}
\begin{equation}\label{eq28}
E_{\rm g.s}^{\rm QRPA}=E_{\rm
BCS}+\frac{1}{2}\left[\sum_{\nu}\omega_{\nu}-\left(\sum_jA_{jj}\right)\right]~.
\end{equation}
\subsubsection{\label{subsec2.3.2}Renormalized QRPA}
To restore the Pauli principle, the QRPA is 
renormalized based on Eq. \eqref{eq14}
instead of the QBA \eqref{eq23}.
The RQRPA operators are introduced as~\cite{QRPA1}
\begin{equation}\label{eq31}
{\cal Q}_\nu^{\dagger}=\sum_j\frac{1}{\sqrt{\langle{\cal D}_{j}\rangle}}\left({\mathcal{X}_j^\nu}\mathcal{A}_j^{\dagger}
-{\mathcal{Y}_j^\nu}\mathcal{A}_j\right)~,
\hspace{5mm} {\cal Q}_\nu=({\cal Q}_\nu^{\dagger})^{\dagger}~,
\end{equation}
which are bosons within the quasiparticle correlated ground state
$|\bar{0}\rangle$, i.e.
\begin{equation}\label{eq32}
\langle\bar{0}|\left[{\cal Q}_\nu ,
{\cal Q}_{\nu'}^{\dagger}\right]|\bar{0}\rangle=\delta_{\nu\nu'}~,
\end{equation}
if the $\mathcal{X}_{j}^{\nu}$ and $\mathcal{Y}_{j}^{\nu}$ amplitudes 
satisfy the same
orthogonality conditions \eqref{eq24}, namely
\begin{equation}\label{ortho}
\sum_j\left({\cal X}_j^{\nu}{\cal X}_j^{\nu'}-
{\cal Y}_j^{\nu}{\cal Y}_j^{\nu'}\right)=\delta_{\nu\nu'}~.
\end{equation}
The RQRPA submatrices are given as
\begin{eqnarray}
A_{jj'}&=&2\left[b_j+2q_{jj'}+2\sum_{j''}q_{jj''}
\bigg(1-\frac{\langle{\cal D}_{j}{\cal D}_{j''}\rangle}
{\langle{\cal D}_{j}\rangle}\bigg)\right]\delta_{jj'}
+d_{jj'}\frac{\langle{\cal D}_{j}{\cal D}_{j'}\rangle}
{\sqrt{\langle{\cal D}_{j}\rangle\langle{\cal D}_{j'}\rangle}}~,
\label{eq33.1}\\
B_{jj'}&=&2h_{jj'}\left(\frac{\langle{\cal D}_{j}
{\cal D}_{j'}\rangle}{\sqrt{\langle{\cal D}_{j}
\rangle\langle{\cal D}_{j'}\rangle}}-\delta_{jj'}\right)~\label{eq33.2}.
\end{eqnarray}
The ground-state correlation factor $\langle{\cal D}_{j}\rangle$ has been derived as a function
of the backward-going amplitudes $\mathcal{Y}_j^\nu$ (see e.g.
Refs. \cite{RRPA5,QRPA1}) as
\begin{equation}\label{eq30}
\langle{\cal D}_{j}\rangle=\frac{1}
{1+\sum_{\nu}\left(\mathcal{Y}_j^\nu\right)^2}~,
\end{equation}
whose values are found by consistently solving Eq. (\ref{eq30}) with 
the RQRPA equations under
the orthogonality condition (\ref{eq24}) for $\mathcal{X}_{j}^{\nu}$ 
and $\mathcal{Y}_{j}^{\nu}$ amplitudes. 
In the limit of $\langle{\cal D}_{j}\rangle=1$, one recovers from 
Eqs. \eqref{eq33.1},
\eqref{eq33.2} the QRPA matrices \eqref{eq27.1} and \eqref{eq27.2}.
\subsubsection{\label{subsec2.3.3}SCQRPA and Lipkin-Nogami SCQRPA }
The only
difference between the SCQRPA and the RQRPA is that, 
similarly to the SCRPA~\cite{SCRPA1,SCRPA2,SCRPA3}, 
the SCQRPA includes the screening factors,
which are the expectation values of the pair operators
$\langle\mathcal{A}_{j'}^{\dagger}\mathcal{A}_j\rangle$ and
$\langle\mathcal{A}_{j'}\mathcal{A}_j\rangle$ over the correlated
quasiparticle ground state $|\bar 0\rangle$. 
The SCQRPA operators are defined in the same way as that for the RQRPA ones so is the
correlated ground state. Therefore we use for it the same
notation $|\bar{0}\rangle$ having in mind the above-mentioned difference
due to screening factors.

The SCQRPA submatrices are obtained in the following form 
\begin{equation}
A_{jj'} = 2\bigg[b_{j}+2q_{jj'}+2\sum_{j''}q_{jj''}
\bigg(1-\frac{\langle{\cal D}_{j}{\cal D}_{j''}\rangle}{\langle
{\cal D}_{j}\rangle}\bigg)
- \frac{1}{\langle{\cal D}_{j}\rangle}\bigg(\sum_{j''}d_{jj''}
\langle{\cal A}_{j''}^{\dagger}{\cal A}_{j}\rangle
- 2\sum_{j''}h_{jj''}\langle{\cal A}_{j''}
{\cal A}_j\rangle\bigg)\bigg]\delta_{jj'}
\label{eq34.1}
\end{equation}
\[
+ d_{jj'}\frac{\langle{\cal D}_{j}
{\cal D}_{j'}\rangle}{\sqrt{\langle{\cal D}_{j}\rangle\langle{\cal D}_{j'}\rangle}}
+ 8q_{jj'}\frac{\langle{\cal A}_{j}^{\dagger}{\cal A}_{j'}\rangle}
{\sqrt{\langle{\cal D}_{j}\rangle\langle{\cal D}_{j'}\rangle}}~,
\]
\[
B_{jj' }= -2\bigg[h_{jj'}+
\frac{1}{\langle{\cal D}_{j}\rangle}\bigg(\sum_{j''}d_{jj''}
\langle{\cal A}_{j''}{\cal A}_{j}\rangle
+ 2\sum_{j''}h_{jj''}
\langle{\cal A}_{j''}
^{\dagger}{\cal A}_{j}\rangle\bigg)\bigg]\delta_{jj'}
\]
\begin{equation}
    + 2h_{jj'}\frac{\langle{\cal D}_{j}
{\cal D}_{j'}\rangle}{\sqrt{\langle{\cal D}_{j}\rangle
    \langle{\cal D}_{j'}\rangle}}
+ 8q_{jj'}\frac{\langle{\cal A}_{j}{\cal A}_{j'}
\rangle}{\sqrt{\langle{\cal D}_{j}\rangle\langle{\cal D}_{j'}\rangle}}
\label{eq34.2}~,
\end{equation}
where the screening factors
$\langle\mathcal{A}_j^{\dagger}\mathcal{A}_{j'}\rangle$ and
$\langle\mathcal{A}_j\mathcal{A}_{j'}\rangle$ are given in terms of
the amplitudes $\mathcal{X}_j^{\nu}$ and
$\mathcal{Y}_j^{\nu}$ as
\begin{eqnarray}
\langle\mathcal{A}_j^{\dagger}\mathcal{A}_{j'}\rangle\equiv
\langle\bar{0}|\mathcal{A}_j^{\dagger}\mathcal{A}_{j'}|\bar{0}\rangle&=&
\sqrt{\langle{\cal D}_{j}\rangle\langle{\cal D}_{j'}\rangle}\sum_{\nu}{\mathcal{Y}_j^{\nu}\mathcal{Y}_{j'}^{\nu}}~,
\label{eq35.1}\\
\langle\mathcal{A}_j\mathcal{A}_{j'}\rangle
\equiv\langle\bar{0}|\mathcal{A}_j\mathcal{A}_{j'}|\bar{0}\rangle
&=&\sqrt{ \langle{\cal D}_{j}\rangle
\langle{\cal D}_{j'}\rangle}\sum_{\nu}{\mathcal{X}_j^{\nu}\mathcal{Y}_{j'}^{\nu}}~.\label{eq35.2}
\end{eqnarray}
The rhs of Eqs. (\ref{eq35.1}) and (\ref{eq35.2}) are obtained by using the
inverted transformation of Eq. (\ref{eq31}), namely
\begin{equation}
{\cal A}_j^{\dagger}=\sqrt{\langle{\cal D}_{j}\rangle}\sum_\nu\left
({\mathcal{X}_j^\nu}\mathcal{Q}_\nu^{\dagger}
+{\mathcal{Y}_j^\nu}\mathcal{Q}_\nu\right)~,
\end{equation}
and Eq. (\ref{eq32}).

For the internal (ground-state) energy, the relation \eqref{eq28} no longer
holds due to the presence of the ground-state correlation factors $\langle{\cal D}_{j}\rangle$
in the SCQRPA equations. Therefore, the SCQRPA ground-state energy is 
calculated directly as the expectation value of the
Hamiltonian \eqref{eq5} in the correlated quasiparticle ground state, 
namely
\[
E_{\rm g.s.}^{\rm SCQRPA}=\langle\bar{0}|H|\bar{0}\rangle
=a+\sum_jb_j(1-\langle{\cal D}_{j}\rangle)+\sum_{jj'}d_{jj'}\langle\mathcal{A}_j^{\dagger}\mathcal{A}_{j'}\rangle
\]
\begin{equation}
+\sum_{jj'}h_{jj'}\left(\langle\mathcal{A}_j^{\dagger}\mathcal{A}_{j'}^{\dagger}\rangle+\langle
\mathcal{A}_{j'}\mathcal{A}_j\rangle\right)
+\sum_{jj'}q_{jj'}\langle\left(1-{\cal D}_{j}\right)
    \left(1-{\cal D}_{j'}\right)\rangle~,\label{eq36}
\end{equation}
In the numerical calculations in the present paper
the exact ratios 
$\langle{\cal D}_{j}{\cal D}_{j'}\rangle/
    \sqrt{\langle{\cal D}_{j}\rangle\langle{\cal D}_{j'}\rangle}$ in
    the RQRPA and SCQRPA submatrices (\ref{eq33.1}),
(\ref{eq33.2}), (\ref{eq34.1}), and (\ref{eq34.2}) are calculated
within the approximation (\ref{approx1}), whose accuracy within the
SCQRPA is numerically tested in the Appendix \ref{appendix}.

Concerning the SCQRPA 
ground-state energy, by
using Eq. (\ref{DD}) and relation (\ref{deltaN}), 
the last term at the rhs of Eq. (\ref{eq36}) can be approximated  as
\[
    \sum_{jj'}q_{jj'}\langle\left(1-{\cal D}_{j}\right)
    \left(1-{\cal D}_{j'}\right)\rangle
    \simeq \sum_{jj'}q_{jj'}\left(1-\langle{\cal D}_{j}\rangle\right)
    \left(1-\langle{\cal D}_{j'}\rangle\right)+\sum_{j}q_{jj}(\delta{\cal
    N}_{j})^{2}
\]
    \begin{equation}
        =-\frac{G\Delta}{4}
    \bigg[\sum_{jj'}\frac{\left(1-\langle{\cal D}_{j}\rangle\right)
    \left(1-\langle{\cal D}_{j'}\rangle\right)}{E_{j}E_{j'}}
    +\sum_{j}\frac{1-\langle D_{j}\rangle^{2}}{2E_{j}^{2}}\bigg]~.
    \label{Egsapprox}
    \end{equation}
The set of Eq. (\ref{uv}) (for $u_{j}$ and $v_{j}$) with the
renormalized single-particle energies $\epsilon'_{j}$ (\ref{reeps})
replacing $\epsilon_{j}$, 
Eq. (\ref{eq26}) with submatrices (\ref{eq34.1}), (\ref{eq34.2}), 
and Eq. (\ref{ortho})
(for the amplitudes ${\cal X}_{j}^{\nu}$, ${\cal Y}_{j}^{\nu}$ and energies
$\omega_{\nu}$), together with Eq. (\ref{eq30}) (for the ground-state
correlation factors $\langle{\cal D}_{j}\rangle$) forms a set of
coupled non-linear equations for $u_{j}$, $v_{j}$, ${\cal X}_{j}^{\nu}$, 
${\cal Y}_{j}^{\nu}$, $\omega_{\nu}$, and $\langle{\cal D}_{j}\rangle$. 
This set is solved
by iteration in the present paper to ensure the self-consistency with
the SCQRPA. Neglecting the screening factors
(\ref{eq35.1}) and (\ref{eq35.2}) the SCQRPA is reduced to the RQRPA,
and the SCQRPA correlated ground state $|\bar{0}\rangle$ becomes the
RQRPA ground state. 

The Lipkin-Nogami SCQRPA (LNSCQRPA) equations have the same form as
that of the SCQRPA ones given in Eqs. \eqref{eq34.1} and
\eqref{eq34.2}, but the chemical potential and  coefficients of the Bogoliubov
transformation are determined by solving the LN1 gap
equations \eqref{eq17.1}, \eqref{eq17.2} instead of the BCS ones.
\section{\label{sec3} Analysis of numerical calculations}
We carried out the calculations of the ground-state energy, $E_{\rm g.s}$,
and energies of excited states, $\omega_{\nu}\equiv {\cal
E}_{\nu}-{\cal E}_{0}$~, in the quasiparticle representation using the 
BCS, QRPA, SCQRPA as
well as their renormalized and PNP versions, 
namely the RBCS, BCS1, LN, RLN, LN1,  
LNQRPA, and LNSCQRPA, at several values of particle
number $N$. The detailed discussion is given for the case with $N=$
10. In the end of the discussion we report a comparison between
results obtained for $N=$ 4, 6, 8, and 10 to see the systematic with
increasing $N$.
\subsection{\label{subsec3.1}Pairing gap}
\begin{figure}
\includegraphics{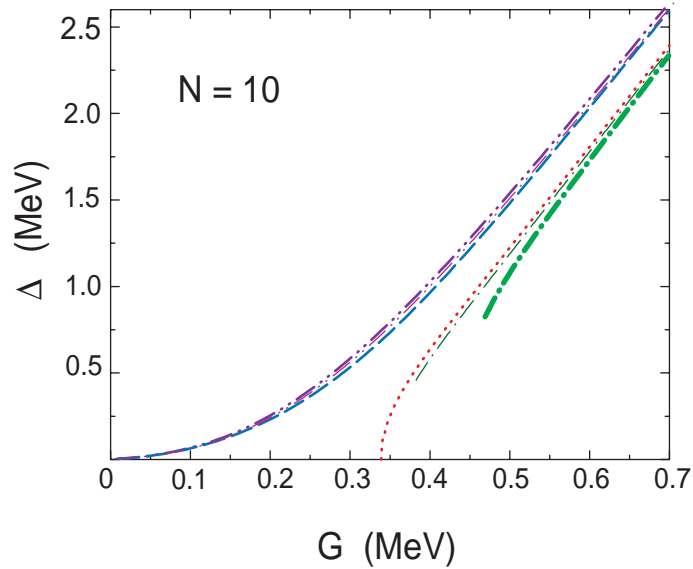}
\caption{(Color on line) Pairing gaps $\Delta$ as functions of $G$ for $N$=10. The
dotted, thin and thick dash-dotted denote the BCS, RBCS, and
BCS1 results, respectively while the dashed, thin and thick 
dash-double-dotted lines represent the
LN, RLN, and LN1 results, respectively.\label{GapN10}}
\end{figure}
Shown in Fig. \ref{GapN10} are the pairing gaps obtained within the BCS, 
RBCS, BCS1,
LN, RLN, and LN1 as functions of the pairing-interaction
parameter $G$ for $N=$ 10. Similarly to the two-level case
\cite{QRPA1}, the BCS has only a trivial solution
$\Delta_{\rm BCS}=$ 0 at $G\leq G_{\rm cr}^{\rm BCS}=$ 0.34 MeV,
while at $G > G_{\rm cr}^{\rm BCS}$ the gap $\Delta_{\rm BCS}$
increases with $G$. Within the BCS1 (RBCS) the ground-state
correlation factor $\langle{\cal D}_{j}\rangle$ 
is always smaller than 1 (at $G\neq$ 0).
This shifts up the value of the critical point $G_{\rm cr}$ to
$G_{\rm cr}^{\rm RBCS}\simeq$ 0.38 MeV, and 
$G_{\rm cr}^{\rm BCS1}\simeq$ 0.47 MeV so that 
$G_{\rm cr}^{\rm BCS}<G_{\rm cr}^{\rm RBCS}<G_{\rm cr}^{\rm BCS1}$.
The PNP within the LN method completely smears out the BCS and
BCS1 (RBCS) critical points to produce the pairing gap $\Delta_{\rm LN}$ as a
smooth function of $G$, which increases with $G$ starting from its
zero value at $G=$ 0. It is worth noticing that, while the 
BCS1 and RLN gaps are
smaller than the BCS ones at a given $G$, especially for the
BCS1 gap at
$G\simeq G_{\rm cr}^{\rm BCS1}$,
the increases of the gap offered by the LN1 and RLN 
compared to the LN value are negligible at all $G$. 
\subsection{\label{subsec3.2}Ground-state energy}
\begin{figure}
\includegraphics{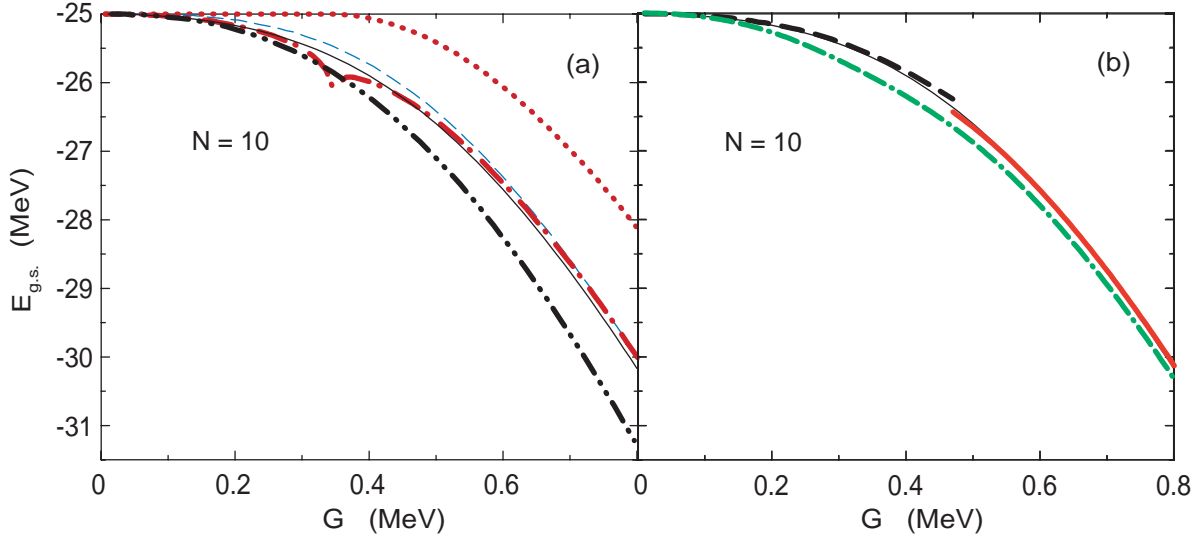}
\caption{(Color on line) Ground state energies as functions of $G$ for $N=10$.
The exact result is represented by the thin solid line in both panels (a) and
(b). In panel (a), the dotted line denotes the BCS
result, the thin dashed line stands for the LN result, 
the dash-dotted line shows the $pp$ RPA result at $G\leq G_{\rm
cr}^{\rm BCS}$, and the QRPA one at $G>G_{\rm cr}^{\rm BCS}$~, while 
the dash--double-dotted line depicts the LNQRPA result. 
Predictions by self-consistent approaches are plotted in panel (b), where 
the thick dashed line denotes the SCRPA result, while 
the SCQRPA and LNSCQRPA are shown by the thick solid and
double-dash--dotted lines, respectively.
\label{EgsN10}}
\end{figure}
Shown in Fig. \ref{EgsN10} are the results for the ground-state energies
obtained within the BCS, LN, SCRPA,
QRPA, LNQRPA, SCQRPA, and LNSCQRPA in comparison with the exact
one for $N=$ 10. The exact result is obtained by directly
diagonalizing the Hamiltonian in the Fock space
\cite{exact2}. It is seen that the BCS strongly overestimates
the exact solution. The LN result comes much closer to the exact one
even in the vicinity of the BCS (QRPA) critical point, while  
the QRPA (RPA) result agrees well with the 
exact solution only at $G\gg G_{\rm cr}^{\rm BCS}$ ($G\ll G_{\rm
cr}^{\rm BCS}$). 
The improvement given by the SCRPA is significant as its result nearly coincides with the exact one
in the weak coupling region. However the convergence of the SCRPA solution is
getting poor in the strong coupling region. As a result, only the values up to 
$G\leq$ 0.46 MeV are accessible. The SCQRPA is much better
than the QRPA as it fits well the exact ground-state energy at $G\geq 
G_{\rm cr}^{\rm BCS1}$. The LNQRPA strongly underestimates 
the exact solution while the LNSCQRPA, which includes the
effects due to the screening factors in combination with PNP, significantly
improves the overall fit. From this analysis, we can say that, among
all the approximations undergoing the test to describe simultaneously the ground and
excited states, the SCRPA, SCQRPA, and
LNSQRPA can be selected as those which fit best the exact ground-state 
energy. The LN method based on the BCS (thin dashed line) also fits quite well the exact one 
at all $G$ 
but it does not allow to describe the excited states as the approaches
based on the QRPA do. Although the fit offered by the LNSCQRPA in the
vicinity of the critical point is somewhat poorer than 
those given by the SCRPA and the
SCQRPA, its advantage is that it does not suffer
any phase-transition point due to the violation of particle number as 
well as the Pauli principle.

\begin{table}
\begin{center} 
    \caption{The energy difference $\Delta E\equiv E_{\rm g.s.}(G)-E_{\rm
    g.s.}(0)$ at various $G$ (in MeV)
    as predicted by the QRPA, SCQRPA, LNQRPA, LNSCQRPA, and exact
    solutions for $N=$ 10.\label{table1}}
    \vspace{2mm}
    
    \begin{tabular}{|c|cccccc|}
    \hline\hline ~~~G~~~ &{~~~QRPA~~~}&
    {~~~SCQRPA~~~}&{~~~LNQRPA~~~}&
    {~~~LNSCQRPA~~~}&{~~~Exact~~~}&
    \\ \hline
    0.10&&&-0.05&-0.06&-0.04& \\
    0.20&&&-0.24&-0.28&-0.17& \\
    0.30&&&-0.63&-0.69&-0.44 &\\
    0.35&-0.93&&-0.91&-0.94&-0.64& \\
    0.40&-1.00&&-1.26&-1.21&-0.90& \\
    0.47&-1.38&-1.44&-1.86&-1.66&-1.36& \\
    0.50&-1.60&-1.66&-2.16&-1.88&-1.60& \\
    0.60&-2.53&-2.58&-3.34&-2.80&-2.56& \\
    0.70&-3.70&-3.75&-4.76&-3.96&-3.76& \\
    0.80&-5.09&-5.13&-6.39&-5.33&-5.17& \\
    0.90&-6.65&-6.68&-8.19&-6.87&-6.75& \\
    1.00&-8.34&-8.38&-10.13&-8.56&-8.46& \\
    1.10&-10.15&-10.18&-12.19&-10.37&-10.29& \\
    1.20&-12.05&-12.08&-14.33&-12.27&-12.22& \\
    1.30&-14.03&-14.06&-16.55&-14.25&-14.22& \\
    1.40&-16.06&-16.10&-18.84&-16.30&-16.28& \\
    \hline\hline
    \end{tabular}
\end{center}
\end{table}
The corrections due to ground-state
correlations can also be clearly seen by examining the energy difference
\begin{equation}
\Delta E\equiv E_{\rm
g.s.}(G)-E_{\rm g.s.}(0)
\label{DeltaE}
\end{equation}
between the ground-state energies defined at
finite and zero $G$~\footnote{Within the RPA and SCRPA, where the 
mean field is the HF one, $\Delta E$ coincides with the
correlation energy $E_{\rm corr}\equiv E_{\rm g.s.} -
E_{\rm HF}$ because
$E_{\rm g.s.}^{(\rm exact)}(0) = E_{\rm HF}$, 
($f^{\rm HF}_{p}=$ 0, $f^{\rm HF}_{h}=$ 1). 
Within the quasiparticle
formalism, however, $E_{\rm corr}$ is defined as the difference
between the QRPA (LNQRPA, SCQRPA, LNSQRPA) 
ground-state energy and that given within the BCS
(LN, LN1) method. This $E_{\rm corr}$ is quite different from 
$\Delta E$ in the strong-coupling regime because of the 
large pairing gap. Therefore
we find more appropriate in the quasiparticle
representation to compare 
the approximated and exact energies $\Delta E$ (\ref{DeltaE}) 
rather than $E_{\rm corr}$.}. The values of this energy difference 
as predicted by the QRPA, SCQRPA, LNQRPA,
and LNSCQRPA for the system with $N=$ 10 at various $G$ are compared with the exact 
ones in Table \ref{table1}. It is seen from this table that, while in the
weak coupling regime ($G_{\rm cr}^{\rm BCS}\leq G\leq$ 0.8 MeV) 
the QRPA and SCQRPA predictions for this energy difference
are closer to the exact result, at high $G$ the SCQRPA and LNSCQRPA
are the ones
that offer the better fits for this quantity. 
The LNQRPA, on the contrary, offers a quite 
poor fit for $\Delta E$ to the exact result.

\begin{table}
\begin{center}
	\caption{Relative errors $\delta E^{(\rm a)}$ and $\delta 
	E^{(\rm b)}$ from Eq. (\ref{err}) at various $G$ 
	as predicted by the QRPA, SCQRPA, LNQRPA, and LNSCQRPA for
	$N=$ 10. \label{table2}}
    \begin{tabular}{|c|cccc|cccc|}
    \hline\hline
     & \multicolumn{4}{c|}{$\delta E^{(\rm a)}$~~($\%$)} & 
    \multicolumn{4}{c|}{$\delta E^{(\rm b)}$
    ~~($\%$)}\\
    \hline
    ~G~~(MeV)& \multicolumn{1}{c}{~~~\scriptsize QRPA}&
    \multicolumn{1}{c}{\scriptsize
    SCQRPA}&\multicolumn{1}{c}{\scriptsize LNQRPA}&
    \multicolumn{1}{c|}{\scriptsize LNSCQRPA}&
    \multicolumn{1}{c}{~~~\scriptsize QRPA}&
    \multicolumn{1}{c}{\scriptsize
    SCQRPA}&\multicolumn{1}{c}{\scriptsize LNQRPA}&
    \multicolumn{1}{c|}{\scriptsize LNSCQRPA}
    \\ \hline
0.10&&&25.00&50.00&&&0.04&0.08 \\
0.20&&&41.18&64.71&&&0.28&0.44 \\
0.30&&&43.18&56.82&&&0.75&0.98 \\
0.35&43.51&&42.19&46.88&1.13&&1.05&1.17 \\
0.40&11.11&&40.00&34.44&0.39&&1.39&1.20 \\
0.47&1.47&5.88&36.76&22.06&0.08&0.30&1.90&1.14 \\
0.50&0.00&3.75&35.00&17.50&0.00&0.23&2.11&1.05 \\
0.60&1.17&0.78&30.47&9.37&0.11&0.07&2.83&0.87 \\
0.70&1.60&0.27&26.60&5.32&0.21&0.03&3.48&0.70 \\
0.80&1.55&0.77&23.60&3.09&0.27&0.13&4.04&0.53 \\
0.90&1.48&1.04&21.33&1.78&0.32&0.22&4.54&0.38 \\
1.00&1.42&0.95&19.74&1.18&0.36&0.24&4.99&0.30 \\
1.10&1.36&1.07&18.46&0.78&0.40&0.31&5.38&0.23 \\
1.20&1.39&1.15&17.27&0.41&0.46&0.38&5.67&0.13 \\
1.30&1.34&1.13&16.39&0.21&0.48&0.41&5.94&0.08 \\
1.40&1.35&1.11&15.72&0.12&0.53&0.44&6.20&0.05 \\
\hline\hline
\end{tabular}
\end{center}
\end{table}
A more quantitative calibrations can be seen by analyzing the relative
errors 
\begin{equation}
\delta E^{(\rm a)} = \frac{\Delta E^{(\rm approx)}-\Delta E^{(\rm exact)}}
{\Delta E^{(\rm exact)}}~,\hspace{2mm} {\rm and}\hspace{ 5mm}    
\delta E^{(\rm b)} = \frac{E^{(\rm approx)}-E^{(\rm exact)}}
{E^{(\rm exact)}}~,
\label{err}
\end{equation}
which are shown in Table \ref{table2}. Because $\Delta E^{(\rm exact)}$ 
are quite small at small $G$, the relative errors $\delta E^{(\rm a)}$
are quite large in the weak-coupling region. In this respect the
relative error $\delta E^{(\rm b)}$ turns out to be a better
calibration. While  $\delta E^{(\rm a)}$ decreases as $G$ increases 
for all approximations with the LNSCQRPA having the smallest relative errors
at large $G$, the behavior of $\delta E^{(\rm b)}$ on $G$ is
somewhat different depending on the approximation. A decrease of this
quantity is seen within the QRPA and SCQRPA with increasing $G$ up to
$G=$ 0.7 MeV, and an increase with $G$ takes place at large $G$. For
the LNSCQRPA, the relative error $\delta E^{(\rm b)}$ increases first
with $G$ up to $G=$ 0.4 MeV, then decreases at larger $G$. Within LNQRPA
one sees a steady increase of $\delta E^{(\rm b)}$ with $G$ to reach a 
value as large as 6.2 $\%$ at $G=$ 1.4 MeV.

\begin{figure}
\includegraphics{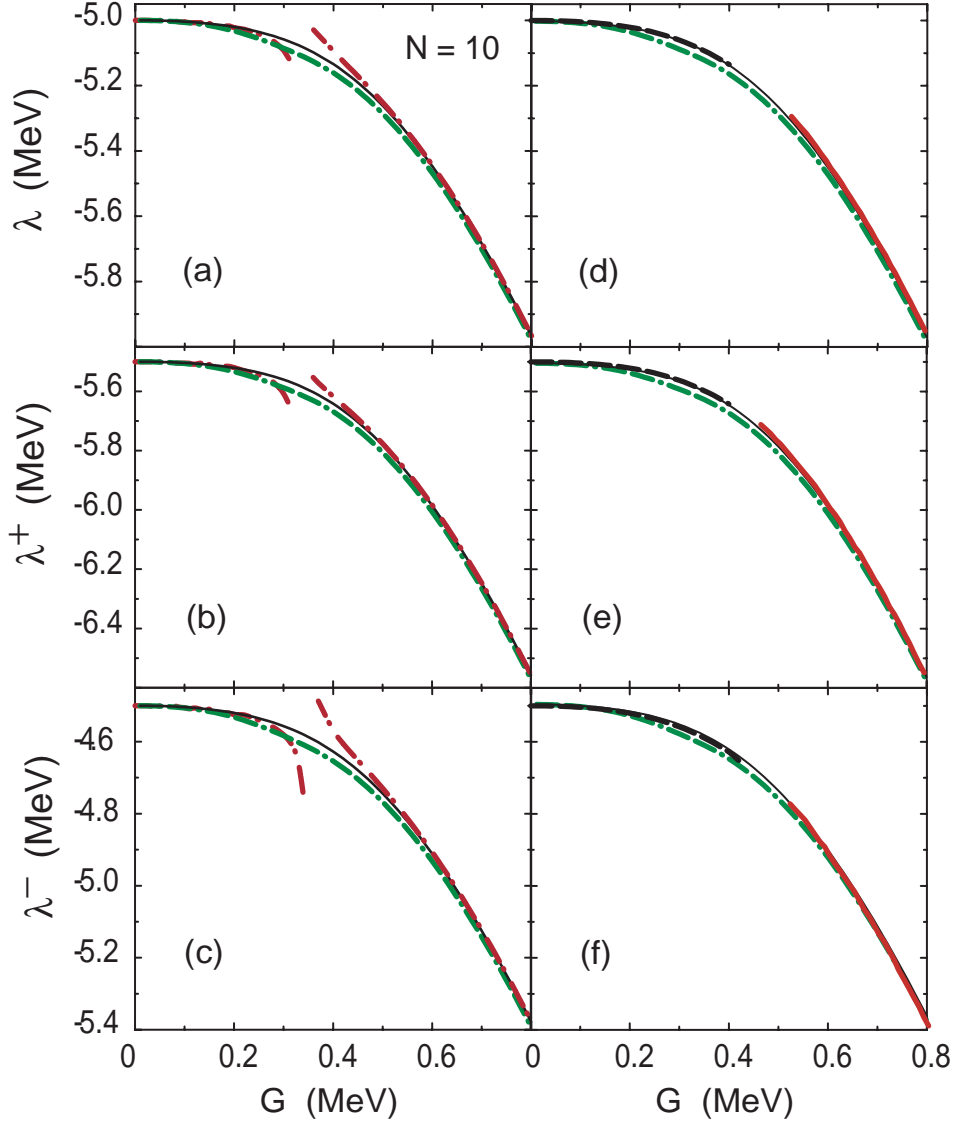}
\caption{(Color online) 
Chemical potentials $\lambda$ and $\lambda^{\pm}$ as functions
of $G$ for $N=$ 10 as predicted by the exact solutions, RPA, QRPA,
SCRPA, SCQRPA, and LNSCQRPA. Notations are as in Fig. \ref{EgsN10}.
\label{lambdafig}}
\end{figure}
The quantities that are directly defined by the differences of
ground-state energies are the chemical potentials $\lambda^{\pm}$ and 
$\lambda$, namely
\begin{equation}
    \lambda^{+}=\frac{1}{2}\left[E_{\rm g.s.}(N+2) - E_{\rm
    g.s.}(N)\right]~,
    \hspace{5mm} 
    \lambda^{-}=\frac{1}{2}\left[E_{\rm g.s.}(N) - E_{\rm
        g.s.}(N-2)\right]~, 
        \hspace{5mm}
        \lambda =\frac{1}{2}(\lambda^{+}+\lambda^{-})~.
        \label{chem}
        \end{equation}
The exact values of the 
chemical potentials $\lambda$ and $\lambda^{\pm}$
are shown in Fig. \ref{lambdafig} in comparison with
the predictions within quasiparticle presentations for $N=$ 10.
It is seen from this figure that 
the SCRPA and SCQRPA [Fig. \ref{lambdafig} (d) - \ref{lambdafig} (f)] 
offer the best fit to the exact results except that 
the SCRPA poorly converges at $G>$ 0.4 MeV, while 
SCQRPA stops at $G=G_{\rm cr}^{\rm BCS1}$. 
The RPA and QRPA also describe very well the exact results, except 
the values in the critical region, where the RPA and QRPA
diverge. The LNSCQRPA predictions for the chemical potentials 
show smooth functions at all $G$, which fit well the exact results,
including the region around $G_{\rm cr}$, where they slightly
underestimates the exact ones.
\subsection{\label{subsec3.3} Energies of excited state}
\begin{figure}
\includegraphics{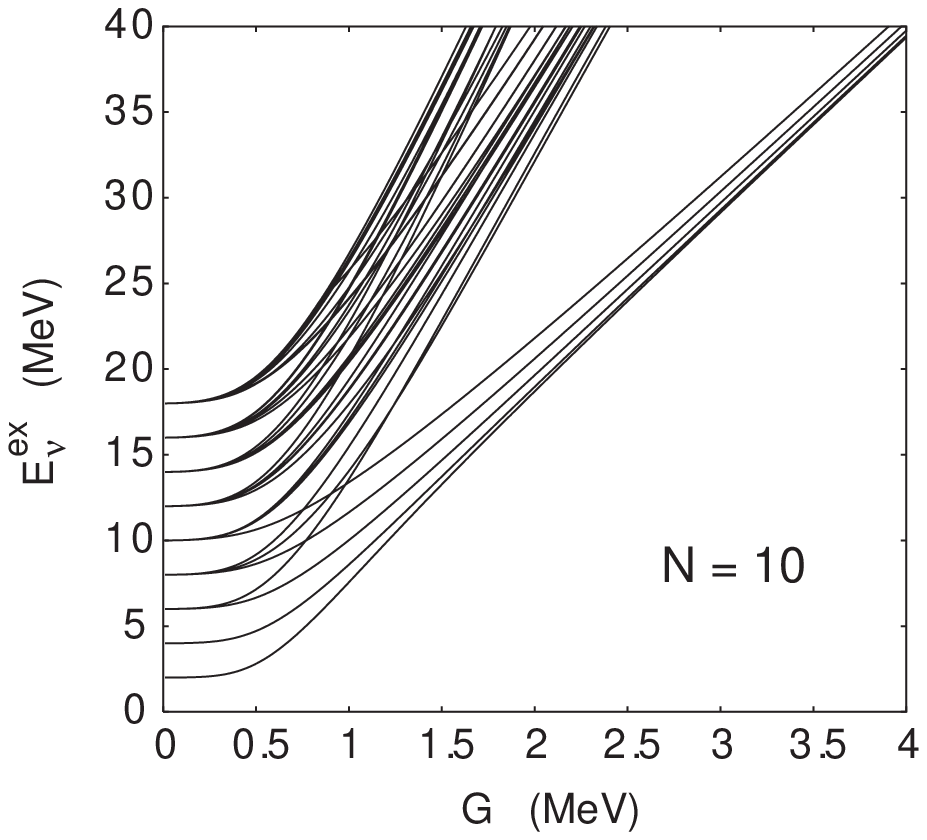}
\caption{Exact energies 
$E^{\rm ex}_{\nu}\equiv{\cal E}_{\nu}^{\rm ex}(N)-{\cal E}_{0}^{\rm ex}(N)$ 
obtained within the Richardson model for
excited states $\nu$ relative to the exact
ground-state level ${\cal E}_{0}^{\rm ex}$ as functions of $G$ for $N$=10.
\label{ExactN10}}
\end{figure}
\begin{figure}
\includegraphics{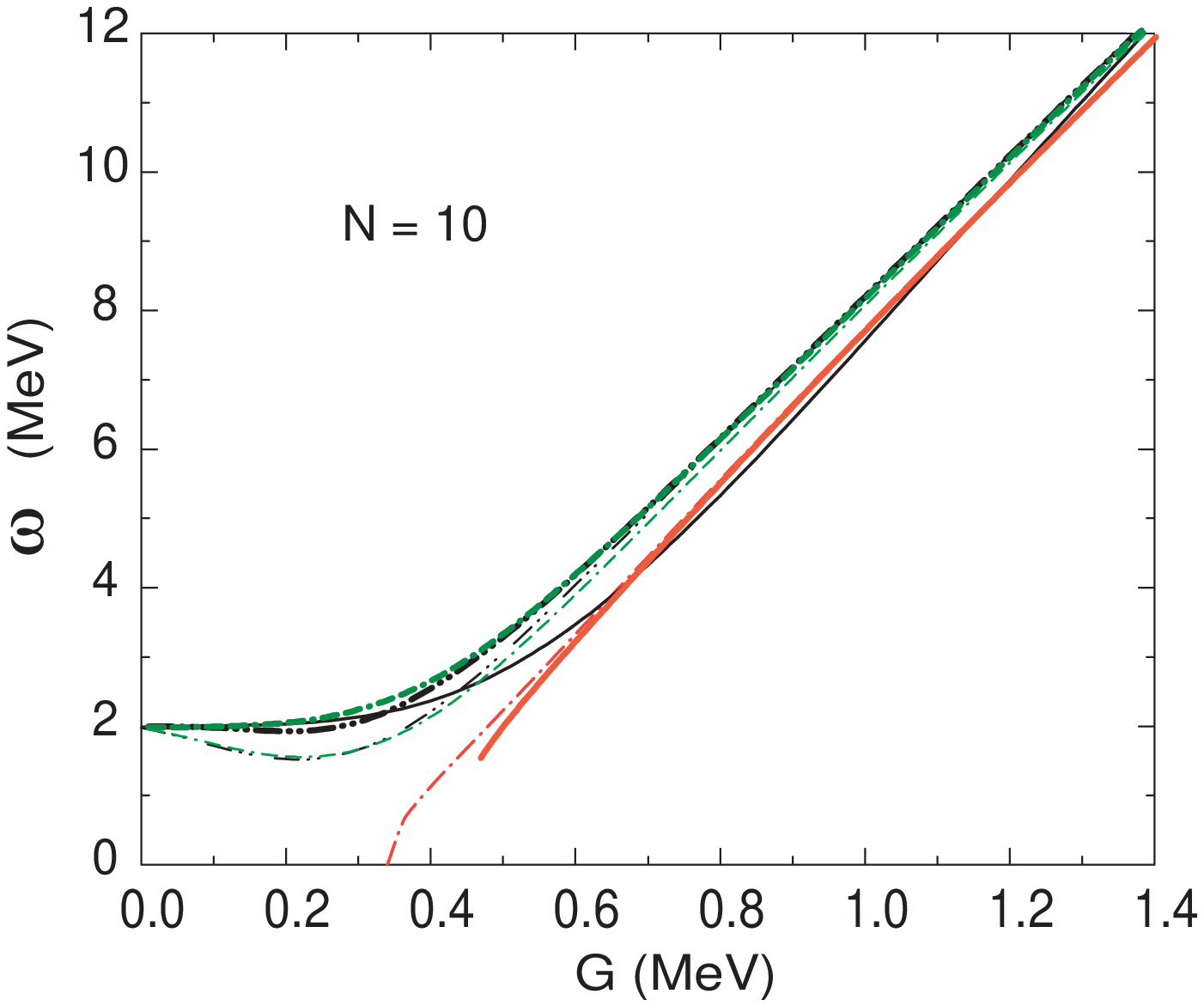}
\caption{(Color on line) The energies of the first excited state as functions of $G$ at
$N$=10. The results refer to the exact solution, $E_{1}^{\rm ex}$ 
(solid line), the QRPA
solution, $\omega_2^{\rm QRPA}$ (dash-dotted line), 
the SCQRPA solution, $\omega_2^{\rm SCQRPA}$ (thick solid line), 
the LNQRPA solutions, $\omega_2^{\rm LNQRPA}$ (thin dash -- double-dotted
line) and $\omega_3^{\rm LNQRPA}$ (thick dash -- double-dotted line),
as well as the LNSCQRPA
solutions, $\omega_2^{\rm LNSCQRPA}$
(thin double-dash -- dotted line) and $\omega_3^{\rm LNSCQRPA}$ (thick double-dash
 -- dotted line).\label{EexN10}}
\end{figure}
As has been discussed in Refs. \cite{QRPA1, SCQRPA2}, the first solution
$\omega_1$ of the QRPA or SCQRPA equations is the energy of spurious
mode, which is well separated from the physical solutions $\omega_\nu$ with $\nu\geq$ 2. 
The first excited state energy is therefore given by
$\omega_2$. Figure \ref{ExactN10} shows the exact eigenvalues for the
excited states. As has also been demonstrated in Ref. \cite{exact3},
this figure shows that the coupling in the small-G region causes only small perturbations in 
the single-particle levels. With increasing $G$ the system goes to the
crossover regime, where level splitting and crossing are seen,
releasing the levels' degeneracy. In the 
strong coupling regime the levels coalesce into narrow well-separated 
bands. 
The approaches based on the QRPA with PNP within the LN method 
also splits the levels but the nature of the splitting comes from
the two components within the
QRPA operator (\ref{eq20}), which correspond to the addition and removal
modes, respectively, in the RPA limit. When the pairing gap $\Delta$
is finite, it is not possible to consider the QRPA excitations as
purely addition or removal modes, but only as those with some components having
the dominating property inherent to one of these modes. 
The QRPA eigenvalues also have two branches 
with positive $\omega_{\nu}$ and negative $-\omega_{\nu}$ energies.
However, unlike the $pp$ RPA, where 
the negative eigenvalues in the
equations for addition modes are also physical as they are 
the energies of the removal modes 
taken with the minus sign and vice versa, within the QRPA only the positive  
energies $\omega_{\nu}$ are physical, and they are compared  with the exact ones, 
$E_{\nu}^{ex}\equiv{\cal E}_{\nu}(N)-{\cal E}_{0}(N)$, in the 
present paper.

\begin{figure}
\includegraphics[width=10cm]{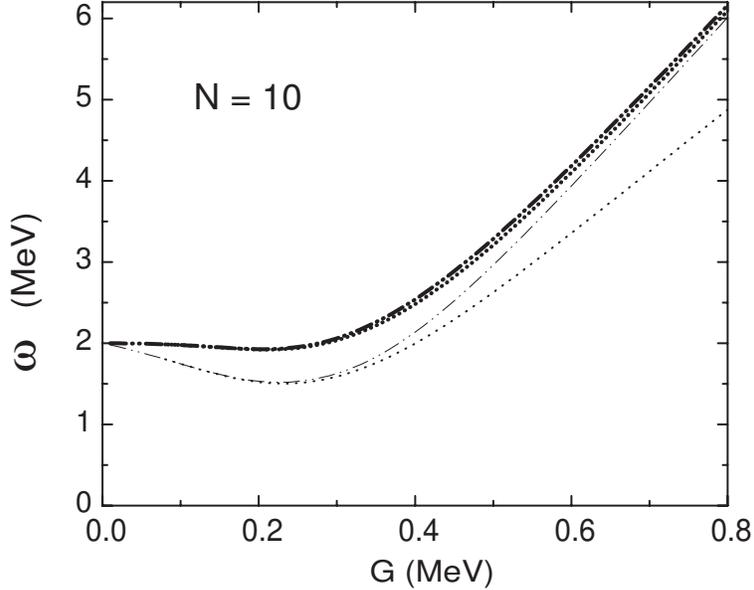}
\caption{The energies of the first excited state in different schemes 
as functions of $G$ for $N=$ 10. The thin and thick dash -- double-dotted
lines denote the second and third LNQRPA solutions, 
while the thin
and thick dotted lines stand for the absolute values of the 
corresponding solutions within the LNQRPA1 scheme.\label{ppLNQRPA}}
\end{figure}

As an example to illustrate this level-splitting pattern, we
show in Fig. \ref{EexN10} the exact energy 
$E_{1}^{\rm ex}\equiv{\cal E}_{1}(N)-{\cal E}_{0}(N)$ 
of the lowest excited state ($\nu=$ 1) with respect to the exact ground
state ($\nu =$ 0) 
in the system with $N=$ 10 particles in comparison with the predictions within 
the QRPA, LNQRPA, SCQRPA, and LNSCQRPA
~\footnote{For the two-level case $E_{1}^{\rm ex}$ corresponds to the
solid line in the upper panel of Figs. 1, 3 -- 5 in Ref.
\cite{SCQRPA2} or Figs. 1 -- 3 in Ref. \cite{Samba} for $N=$ 4, 8, and
12).}. As the exact energy $E_{1}^{\rm ex}$ 
represents the energy of the lowest pair-vibration state,
it is compared with the energies $\omega_{1}$ of the lowest excited state
obtained within QRPA, LNQRPA, SCQRPA and LNSCQRPA, which are built
on the pairing condensate (quasiparticle vacuum). 
The splitting is clearly seen from Fig. \ref{EexN10} within the LN method,
namely the LNQRPA and LNSCQRPA. One can see that, 
within the LN(SC)QRPA, each single level at $G=$ 0 splits 
into two components in the small-G 
region, e.g. the pair $\omega_2^{\rm LNQRPA}$ and $\omega_3^{\rm LNQRPA}$ 
or $\omega_2^{\rm LNSCQRPA}$ and $\omega_3^{\rm LNSCQRPA}$. 
To look inside the source of the splitting, we rewrite the QRPA operator 
(\ref{eq20}) into the components
with dominating contributions of addition- and removal-mode patterns as follows:
\[
Q_\nu^{\dagger}=(Q_\nu^{\dagger})^{(\rm A)}+(Q_\nu^{\dagger})^{(\rm R)}~,
\hspace{5mm} 
\]
\begin{equation}\label{eq37}
    (Q_\nu^{\dagger})^{(\rm A)}=\sum_pX_p^\nu\mathcal{A}_p^{\dagger}
-\sum_hY_h^\nu\mathcal{A}_h~,\hspace{5mm} 
(Q_\nu^{\dagger})^{(\rm R)}=\sum_hX_h^\nu\mathcal{A}_h^{\dagger}
-\sum_pY_p^\nu\mathcal{A}_p~,
\end{equation}
where the indices $j$ run over all the levels, from which those 
located below (above) the chemical potential are formally labelled with
$h$ ($p$) indices. It is not difficult to see that, 
in the RPA limit (or zero-pairing limit),  $(Q_\nu^{\dagger})^{(\rm A)}$ is transformed into
operator $A_\nu^{\dagger}$ that generates the addition modes, while
$(Q_\nu^{\dagger})^{(\rm R)}$ becomes $R_\nu^{\dagger}$ that generates the removal
modes (in the standard notations for addition and
removal operators from Refs.~\cite{SCRPA1,SCRPA2,SCRPA3}). 
Using this formal expression \eqref{eq37}, 
we derived the QRPA equations for the
excitations generated by operators 
$(Q_\nu^{\dagger})^{(\rm A)}$ and $(Q_\nu^{\dagger})^{(\rm R)}$,
separately. The energies of the corresponding 
first excited states from the resulting sets of equations 
were calculated by using the LN method. 
We call this scheme as LNQRPA1. The set of equations for 
the modes generated by operator $(Q_\nu^{\dagger})^{(\rm A)}$ gives a negative 
$\omega_2^{\rm LNQRPA1}$ and positive $\omega_3^{\rm LNQRPA1}$, which 
means that they correspond to the energies of the removal and addition 
modes, respectively. The absolute values of these 
energies are shown in Fig. \ref{ppLNQRPA} 
along with $\omega_{2,3}^{\rm LNQRPA}$. 
It is seen from this figure that in the weak-coupling region 
the higher-lying levels $\omega_3^{\rm LNQRPA}$ and
$\omega_3^{\rm LNQRPA1}$ nearly coincide, 
while the lower-lying one, $\omega_2^{\rm
LNQRPA}$, is almost the same as $|\omega_2^{\rm LNQRPA1}|$. 
From here, we can identify $\omega_3^{\rm LNQRPA}$
and $\omega_2^{\rm LNQRPA}$ as the levels where the addition and
removal modes dominate, respectively. As the interaction 
$G$ increases, the occupation probabilities of the levels below and above the Fermi level
become comparable so it becomes more and more difficult to separate
the patterns belonging to addition and removal modes in the QRPA
excitations. 

From this analysis and Fig. \ref{EexN10}, it becomes clear that, 
in the weak coupling
region, the level $\omega_3^{\rm LNQRPA}$, which is generated mainly
by the addition mode, fits well the exact result, while the
agreement between the exact energy and $\omega_2^{\rm QRPA}$
as well as $\omega_2^{\rm SCQRPA}$ is good only in the strong
coupling region. At large values of $G$, predictions by all
approximations and the exact solution coalesce into one band, whose width 
vanishes in the limit $G\rightarrow\infty$.\\
\begin{figure}
\includegraphics{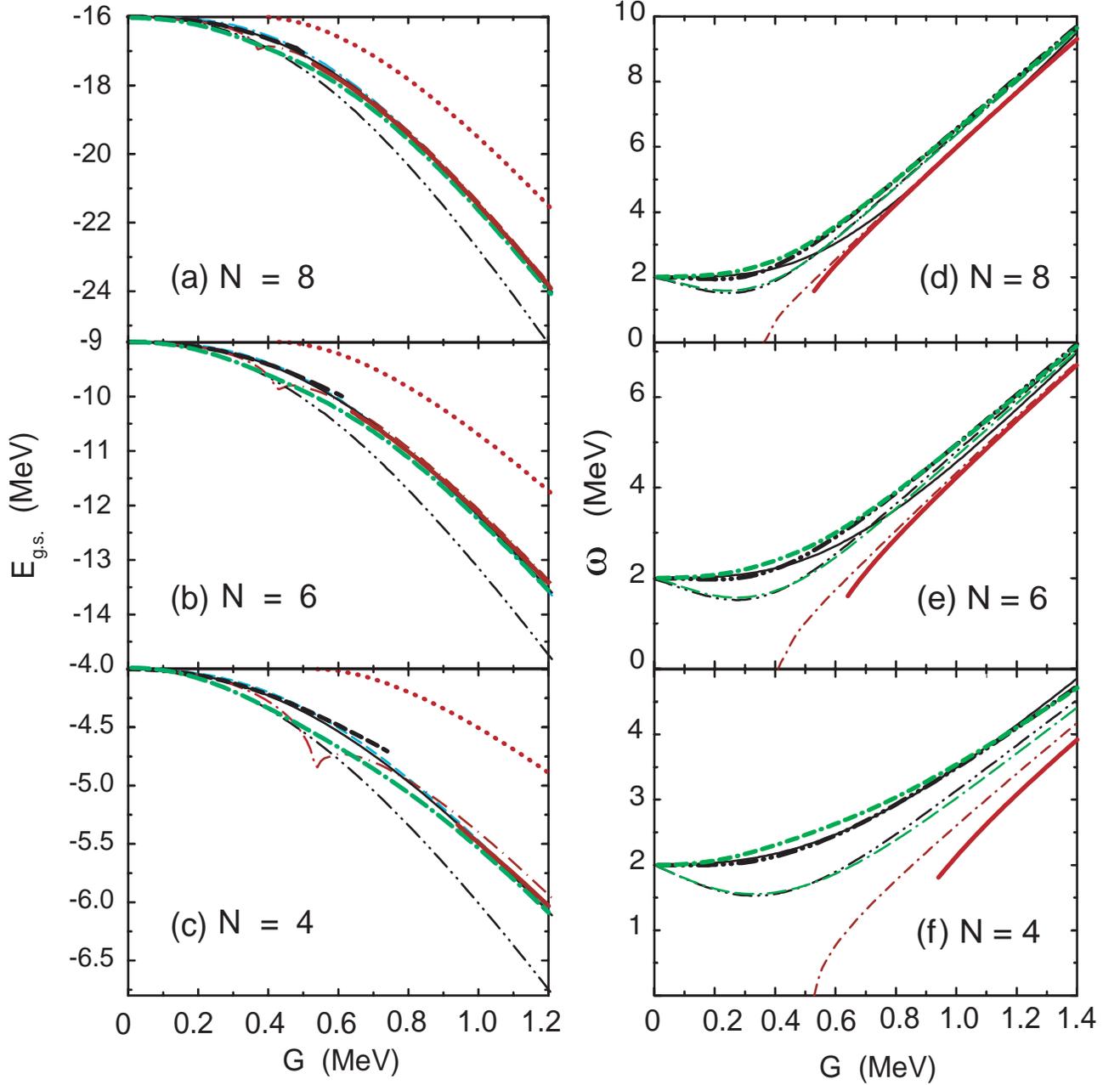}
\caption{(Color on line) Energies of ground state (left panels) (notations as in
Fig. \ref{EgsN10}) and
first excited state (right panels) (notations as in Fig.
\ref{EexN10}) for several values of $N$ indicated on the panels as
functions of $G$.\label{EN468}}
\end{figure}
\indent The energies of the ground state and the first excited state
obtained for $N =$ 4, 6, 8 are depicted in Fig. \ref{EN468}. The
figure shows that increasing $N$ worsens the agreement of the results obtained within
the LNQRPA and LNSCQPPA with the exact ones for both the ground
state and the first excited state, while the QRPA and SCQRPA
results become closer to the exact ones at $G\gg G_{\rm cr}$. At small $N$ ($N=$ 4), the
solution $\omega_{3}^{\rm LNQRPA}$ seems to fit best the exact result for 
all values of $G$.

\begin{figure}
\includegraphics{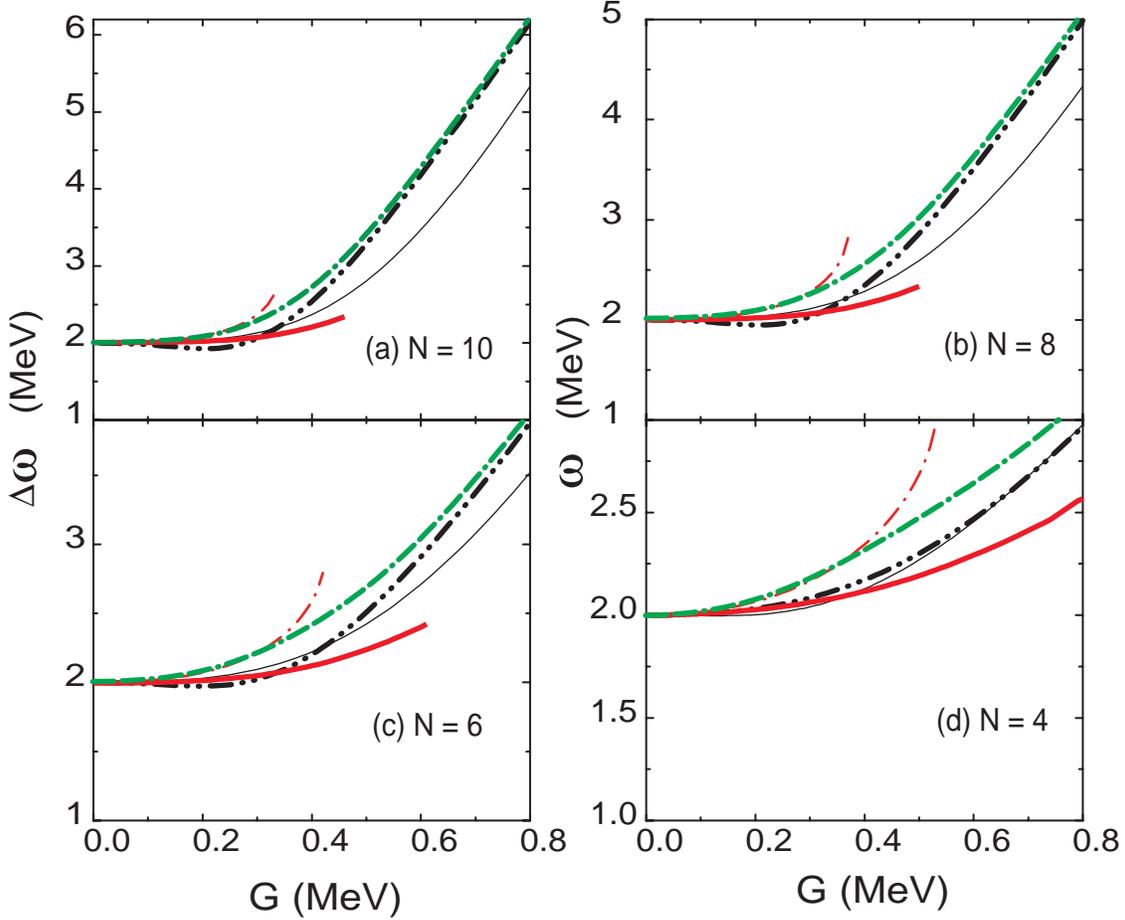}
\caption{(Color on line) Energy $\Delta\omega^{\rm (SC)RPA}$
(\ref{omega1}) obtained within the
$pp$ RPA (dash-dotted line) and SCRPA (thick solid line) 
as a function of $G$ for several values of $N$ in comparison with 
the energy $\omega_{3}^{\rm LNQRPA}$ 
(dash -- double-dotted line), $\omega_{3}^{\rm LNSCQRPA}$ (double-dash
-- dotted line), and 
the exact energy $E_{1}^{\rm ex}$ (thin solid line), which are the
same as those in Fig.
\ref{EN468} (d) -- (f) for $N=$ 4, 6, and 8.
\label{compare}}
\end{figure}

The pair-vibration excitation energy $E_{1}^{\rm ex}$ 
is usually larger than the energy of the 
lowest state with one broken pair.
The latter is
described within the $pp$ RPA as the energy of the lowest addition
mode in the laboratory reference frame fixed to the ground state of
$N$-particle system~\cite{SCRPA1,SCRPA2,SCRPA3}. 
It is worthwhile to compare the
predictions for the excited-state energies 
obtained within the quasiparticle approaches developed in the present 
paper with $pp$ RPA and SCRPA predictions by transforming the
latter into the intrinsic reference frame of the system with $N+2$
particles. This is done as follows. 
From the (SC)RPA energy of the ground-state level ${\omega}_{0}^{\rm
(SC)RPA}={\cal E}_{0}^{\rm (SC)RPA}(N+2) -  {\cal E}_{0}^{\rm (SC)RPA}(N)$, and
that of the first excited state ${\omega}_{1}^{\rm (SC)RPA}=
{\cal E}_{1}^{\rm (SC)RPA}(N+2) -  {\cal E}_{0}^{\rm (SC)RPA}(N)$
\footnote{The energies ${\omega}_{0}^{\rm (SC)RPA}$ and ${\omega}_{1}^{\rm
(SC)RPA}$ correspond to energies $E_{1}$ and $E_{2}$ shown in 
Figs. 3 and 4 in Ref. \cite{SCRPA1}, respectively.} it follows that
\begin{equation}
    \Delta\omega^{\rm (SC)RPA}\equiv 
    {\omega}_{1}^{\rm (SC)RPA}- {\omega}_{0}^{\rm (SC)RPA}
    = {\cal E}_{1}^{\rm (SC)RPA}(N+2) - {\cal E}_{0}^{\rm (SC)RPA}(N+2)~,
    \label{omega1}
    \end{equation}
This energy $\Delta\omega^{\rm (SC)RPA}$ is shown in Fig. \ref{compare}
as a function of $G$ along with the corresponding LNQRPA, LNSCQRPA, and exact
energies for several values of $N$. This figure clearly shows that
the LNQRPA and LNSCQRPA are superior to the $pp$ RPA and SCRPA as they 
offer an overall prediction closer to the exact result for all $G$ and $N$. 
They neither collapse at a $G_{\rm cr}$ as in the case with the $pp$ RPA 
nor have a poor convergence as the SCRPA does at $G \gg G_{\rm cr}$.  
\section{\label{sec4}Conclusions}
This work proposes a self-consistent version of the QRPA in
combination with particle-number projection within the Lipkin-Nogami
method as an approach that works at any values of the pairing-interaction 
parameter $G$ without suffering a phase-transition-like collapse (or
poor convergence) due
to the violation of Pauli principle as well as of the integral of motion 
such as the particle number. 
The self-consistency is maintained within a set of coupled equations
for the pairing gap, QRPA amplitudes, and energies by means
of the screening factors, which are the expectation values of the
products of quasiparticle-pair operators, and the ground-state
correlation factor, which is a function of the QRPA backward-going amplitudes.

The proposed approach 
is tested in a multi-level exactly solvable model, namely the
Richardson model for pairing. The energies of the ground and
first-excited states are calculated within several approximations 
such as the BCS, RBCS, BCS1, LN, RLN, LN1, QRPA, SCQRPA, LNQRPA and
LNSCQRPA. The obtained results for the
ground-state energy show that the  
use of the LN method that includes the SCQRPA correlations not only allows
us to avoid the collapse of the BCS as well as the QRPA but also
fits well the exact result. For the energy of the first excited
state, the LNQRPA and LNSCQRPA results offer the best fits to the exact solutions in the
weak coupling region with large particle numbers, while the QRPA and SCQRPA reproduce well the
exact one in the strong coupling region. In the limit of very large
$G$ all the approximations predict nearly the same value as that of the
exact one.  As the number of particles decreases,
it becomes sufficiently well to use the predictions given by the LNQRPA and LNSCQRPA for 
energies of both the ground state and first-excited state to fit the exact results.

We believe that the approach proposed in this work can be useful in the applications
to light and unstable nuclei, where the validity of the QBA and that
of the conventional BCS are in question. Such applications are 
the goal for forthcoming studies.
\begin{acknowledgments}
The authors are grateful to Michelangelo Sambataro (Catania) 
for his assistance in the exact solutions of the Richardson model.

The numerical calculations were carried out using the FORTRAN IMSL
Library by Visual Numerics on the RIKEN Super Combined Cluster
(RSCC) system. NQH is a RIKEN Asian Program Associate.
\end{acknowledgments}
\appendix
\section{Accuracy of approximation (\ref{approx1})}
\label{appendix}
\begin{table}
\begin{center} 
    \caption
	{BCS1 and LN1 pairing gaps (in MeV) at various values of
	G (in MeV) (see text).\label{tableA1}}
	\vspace{2mm} 
\begin{tabular}{|c|ccc|ccc|}
\hline\hline
& \multicolumn{3}{c|}{BCS1} & \multicolumn{3}{c|}{LN1}\\
\hline G &
\multicolumn{1}{c}{~~~~~$\Delta$~~~~~}&\multicolumn{1}{c}{~~~$\overline{\Delta}$
~~~}&
\multicolumn{1}{c|}{~~~$\frac{\delta\Delta}{\Delta}~(\%)$~~~}&
\multicolumn{1}{c}{~~~~~$\widetilde{\Delta}$~~~~~}&\multicolumn{1}{c}
{~~~$\overline{\widetilde{\Delta}}$~~~}&
\multicolumn{1}{c|}{~~~$\frac{\delta\tilde\Delta}{\tilde\Delta}~(\%)$~~~}\\
\hline 0.01&&&&0.0015&0.0015&0.0000 \\
0.10&&&&0.0606&0.0607&0.1647 \\
0.20&&&&0.2279&0.2289&0.4369 \\
0.30&&&&0.5278&0.5321&0.8081 \\
0.40&&&&0.9579&0.9660&0.8385 \\
0.47&0.8224&0.8357&1.5915&1.3139&1.3233&0.7103 \\
0.50&1.0694&1.0829&1.2467&1.4742&1.4839&0.6537 \\
0.60&1.7219&1.7351&0.7608&2.0261&2.0360&0.4862 \\
0.70&2.3314&2.3436&0.5206&2.5896&2.5993&0.3617 \\
0.80&2.9279&2.9391&0.3811&3.1541&3.1633&0.2908 \\
0.90&3.5132&3.5235&0.2923&3.7148&3.7234&0.2310 \\
1.00&4.0882&4.0977&0.2318&4.2701&4.2783&0.1917 \\
1.10&4.6539&4.6629&0.1930&4.8197&4.8277&0.1657 \\
1.20&5.2118&5.2203&0.1628&5.3641&5.3718&0.1433 \\
1.30&5.7628&5.7710&0.1421&5.9037&5.9113&0.1286 \\
1.40&6.3079&6.3160&0.1282&6.4390&6.4466&0.1179 \\
\hline\hline
\end{tabular} 
\end{center}
\end{table}
\begin{table}
\begin{center} 
    \caption
        {The ratio $(\delta{\cal N}_{j})^{2}/\langle{\cal D}_{j}\rangle$ 
        from Eqs. (\ref{ratio}) and (\ref{deltaN}) 
	   corresponding to the 5 lowest levels $j=$ 1, \ldots, 5, and
	   the energies $\omega_{3}$ (in MeV) of the first excited state
	   described in the text for $N=$ 10
	   at different values of $G$ (in MeV) within the LNSCQRPA. The
	   energy $\omega_{3}(a)$ is obtained including the last term at 
	   the rhs of Eq. (\ref{ratio}), while $\omega_{3}(b)$ is calculated using the
	   approximation (\ref{approx1}).\label{tableA2}}
        \vspace{2mm} 
        
\begin{tabular}{|c|ccccc|ccc|}
\hline\hline ~~~G ~~~~ &~~ $j=1$~~~ &~~ $j=2$~~~ & ~~$j=3$~~~ &
$~~j=4$~~~ & $~~j=5$~~~ & ~~$\omega_{3}(a)$~~~&
~~$\omega_{3}(b)$~~~&\\
\hline
 0.01~~ & 0.0000 & 0.0000 & 0.0000 & 0.0000 & 0.0000 & 2.0001 & 2.0001 & \\
 0.2~~~ & 0.0009 & 0.0012 & 0.0017 & 0.0027 & 0.0046 & 2.0697 & 2.0711 & \\
 0.4~~~ & 0.0023 & 0.0030 & 0.0040 & 0.0055 & 0.0082 & 2.6701 & 2.6742 & \\
 0.6~~~ & 0.0019 & 0.0023 & 0.0027 & 0.0032 & 0.0054 & 4.2040 & 4.2067 & \\
 0.8~~~ & 0.0013 & 0.0015 & 0.0016 & 0.0021 & 0.0033 & 6.1514 & 6.1531 & \\
 1.0~~~ & 0.0009 & 0.0010 & 0.0011 & 0.0015 & 0.0022 & 8.1798 & 8.1812 & \\
 1.2~~~ & 0.0006 & 0.0007 & 0.0009 & 0.0012 & 0.0017 & 10.211 & 10.212 & \\
 1.4~~~ & 0.0005 & 0.0006 & 0.0008 & 0.0011 & 0.0014 & 12.229 & 12.230 & \\
 \hline\hline
\end{tabular} 
\end{center}
\end{table}
Let us analyze the accuracy of the assumption (\ref{approx1})
used in the numerical solutions of the BCS1, LN1, and SCQRPA equations in the
present paper. 

Shown in the 2nd and 5th columns of Table \ref{tableA1} are the values of
the pairing gaps $\Delta$ and $\widetilde\Delta$ 
obtained under the
approximation (\ref{approx1}) within the BCS1 and LN1 method, respectively.
They are compared with the average gaps $\overline{\Delta}$ (3rd column) and
$\overline{\widetilde\Delta}$ (6th column), which are the values obtained by
averaging the level-dependent BCS1 gap $\Delta_{j}$ and LN1 gap
$\widetilde\Delta_{j}$ over all the levels, namely
$\overline{\Delta}=\sum_{j}\Delta_{j}/N$ and 
$\overline{\widetilde\Delta}=\sum_{j}\widetilde\Delta_{j}/N$.
The second term at the rhs of Eq. (\ref{ratio}), which contains
$\delta{\cal N}_{j}^{2}$ as evaluated by the approximation 
(\ref{deltaN}), is taken into account
in calculating $\Delta_{j}$ and $\widetilde\Delta_{j}$ within the
perturbation theory, i.e. with $n_{j}$ being evaluated within SCQRPA 
and LNSCQRPA (where this term is neglected). Except for the two values
at $G=G_{\rm cr}^{\rm BCS1}=$ 0.47 MeV and $G=$ 0.5 MeV within the
BCS1, we see that the values of the relative errors 
$\delta\Delta/\Delta\equiv(\overline\Delta-\Delta)/\Delta$ and 
$\delta\widetilde\Delta/\widetilde\Delta\equiv
(\overline{\widetilde\Delta}-\widetilde\Delta)/\widetilde\Delta$
are all smaller than 1 $\%$, and decrease with increasing $G$.

Shown in Table \ref{tableA2} are the values of
the ratio $(\delta{\cal N}_{j})^{2}/\langle{\cal D}_{j}\rangle$ 
from Eqs. (\ref{ratio}) and (\ref{deltaN}) 
corresponding to the five lowest levels for $N=$ 10 at various $G$
obtained within the LNSCQRPA. 
The largest value of this ratio is observed at the level with $j=$ 5, 
the closest one to the
Fermi level,  at $G=$ 0.4 MeV (close to $G_{\rm cr}^{\rm BCS1}$).
But it amounts to only 0.0082, which is 
a clear evidence that this ratio is indeed negligible.
The last two columns of this table display the  energies
$\omega_{3}(a)$, obtained within the LNSCQRPA including the last term at 
the rhs of Eq. (\ref{ratio}), and $\omega_{3}(b)$, which the LNSCQRPA 
predicts within the approximation (\ref{approx1}). Although a systematic
$\omega_{3}(a)>\omega_{3}(b)$ is observed, the largest difference, 
also seen at $G=$ 0.4 MeV, does not exceed 0.15 $\%$. 
These results guarantee the high accuracy of the
approximation (\ref{approx1}).


\end{document}